\documentclass[12pt,preprint]{aastex}
\shorttitle{Reverberation Radius of Inner Dust Torus}
\shortauthors{Suganuma et al.}

\begin{document}

\title{REVERBERATION MEASUREMENTS OF THE INNER RADIUS OF THE DUST TORUS
IN NEARBY SEYFERT 1 GALAXIES}

\author{Masahiro Suganuma\altaffilmark{1,2},
    Yuzuru Yoshii\altaffilmark{3,4}, Yukiyasu Kobayashi\altaffilmark{1},
    Takeo Minezaki\altaffilmark{3}, Keigo Enya\altaffilmark{5},
    Hiroyuki Tomita\altaffilmark{3},
    Tsutomu Aoki\altaffilmark{7}, Shintaro Koshida\altaffilmark{6,1},
    and
    Bruce A. Peterson\altaffilmark{8}}

\altaffiltext{1}{National Astronomical Observatory,
    2-21-1 Osawa, Mitaka, Tokyo 181-8588, Japan}

\altaffiltext{2}{e-mail: suganuma@merope.mtk.nao.ac.jp}

\altaffiltext{3}{Institute of Astronomy, School of Science, University of
Tokyo, 2-21-1 Osawa, Mitaka, Tokyo 181-0015, Japan}

\altaffiltext{4}{Research Center for the Early Universe,
    School of Science, University of Tokyo,
    7-3-1 Hongo, Bunkyo-ku, Tokyo 113-0033, Japan}

\altaffiltext{5}{Institute of Space and Astronomical Science,
    Japan Aerospace Exploration Agency,
    3-1-1, Yoshinodai, Sagamihara, Kanagawa, 229-8510, Japan}

\altaffiltext{6}{Department of Astronomy, School of Science, University of 
Tokyo, 7-3-1 Hongo, Bunkyo-ku, Tokyo 113-0013, Japan}

\altaffiltext{7}{Kiso Observatory,
    Institute of Astronomy, School of Science, University of Tokyo,
    10762-30 Mitake, Kiso, Nagano 397-0101, Japan}

\altaffiltext{8}{Mount Stromlo Observatory,
    Research School of Astronomy and Astrophysics,
    The Australian National University,
    Weston Creek P. O., A.C.T. 2611, Australia}

\begin{abstract}

The most intense monitoring observations yet made in the optical ($UBV$)
and near-infrared ($JHK$) wave bands were carried out for several 
well-known nearby Seyfert 1 galaxies of NGC 5548, NGC 4051, NGC 3227, 
and NGC 7469.  Over three years of observations, after the operation 
of the MAGNUM telescope started in early 2001, clear time-delayed response of the 
$K$-band flux variations to the $V$-band flux variations was detected 
for all of these galaxies.  Their $H-K$ color temperature was 
estimated to be $1500-1800$ K from the observed flux variation gradients.
This estimated $H-K$ color, along with the smoothness of the flux 
variation in the near-infrared in contrast with rapid fluctuations 
in the optical, supports a view that the bulk of the $K$ flux should 
originate in the thermal radiation of hot dust that surrounds the 
central engine.  Cross-correlation analysis was performed in order 
to quantify the lag time corresponding to the light-travel distance 
of the hot dust region from the central engine.  
The measured lag time is $47-53$ days for NGC 5548, $11-18$ days for 
NGC 4051, about $20$ days for NGC 3227, and $65-87$ days for NGC 7469.
We found that the lag time is tightly correlated with the optical 
luminosity as expected from dust reverberation ($\Delta t \propto 
L^{0.5}$), while only weakly with the central virial mass, 
which suggests that an inner radius of the dust torus around the active 
nucleus has a one-to-one correspondence to central luminosity. In the 
lag time versus central luminosity diagram, the $K$-band lag times 
place an upper boundary on the similar lag times of broad-emission 
lines in the literature.  This not only supports the unified scheme 
of AGNs, but also implies a physical transition from the BLR out to 
the dust torus that encircles the BLR.  
Furthermore, our $V$-band flux variations of NGC 5548 on timescales of up
to 10 days are found to correlate with X-ray variations and delay behind 
them by one or two days, indicating the thermal reprocessing of X-ray 
emission by the central accretion flow. 

\end{abstract}


%%%%%%%%%%%%%%%%%%%%%%%%%%%%%%%%%%%%%%%%%%%%%%%%%%%%%%%%%%%%%%%%%%%%%%%%%%%

\section{INTRODUCTION}

%%%%%%%%%%%%%%%%%%%%%%%%%%%%%%%%%%%%%%%%%%%%%%%%%%%%%%%%%%%%%%%%%%%%%%%%%%%

The unified scheme of active galactic nuclei (AGNs) and radio-quiet quasars 
explains that Seyfert 1 nuclei with broad emission-line regions (BLR) 
would be classified as Seyfert 2 if high velocity clouds emitting broad
emission lines were hidden by the dust tori that encircle such clouds 
(Antonucci 1993).  Outside the dust torus there would be a narrow 
emission-line region (NLR) that extends to more than kilo-parsecs 
and is often observed as resolved images.  However, the angular scale of 
the inner dust torus even in the nearest AGN is too small to be spatially 
resolved with any current technology in optical and near-infrared imaging 
observations.

It is generally accepted that the near-infrared emission from Seyfert 
galaxies and quasars is thermally produced by hot dust heated by UV 
radiation from the central engine.  Barvainis (1987, 1992) developed a 
dust-emission model that naturally explains the bump or excess in the 
near-infrared continuum of AGNs by thermal emission from grains at 
their sublimation temperature in a range of $1300-2000$ K.  This model,
which is supported by the fact that a variety of AGNs have 
blackbody components of nearly uniform temperature of $\sim 1500$ K 
(Kobayashi et al. 1993), necessarily claims a dust distribution 
with central hole.  A typical UV luminosity of 
$10^{42}-10^{44}$ ergs s$^{-1}$ for Seyfert galaxies therefore gives 
an inner radius of $0.01-0.1$ pc or $10-100$ light days at which hot 
dust sublimates.  Consequently, if we monitor their continuum fluxes 
in the UV/optical and near-infrared, a lag time would be expected that 
corresponds to the light-travel distance between the central engine and
the hot dust region from which the bulk of near-infrared emission  
radiates (e.g., Clavel, Wamsteker \& Glass 1989; Nelson 1996; Glass 2004).  

The most successful similar technique is the reverberation mapping 
of the BLR carried out for many Seyfert 1 galaxies and quasars during the 
last decade (e.g., Peterson 1993; Wandel, Peterson \& Malkan 1999; Kaspi 
et al. 2000). If we verify that the infrared lag is longer than the
lag of broad emission lines of lower ionization energy, the inner radius 
of the dust torus should be larger than the outer extension of the BLR, which 
provides strong support to the unified scheme of AGNs.  However, there 
have only been a few cases where lag times for both BLR and dust torus were 
measured for the same AGN (Clavel, Wamstekar \& Glass 1989; Minezaki et 
al. 2004; Suganuma et al. 2004), mostly because long-term simultaneous 
monitoring observations in the UV/optical and near-infrared for dust 
reverberation have rarely been attempted.

The MAGNUM (\underline{M}ulticolor \underline{A}ctive \underline{G}alactic 
\underline{Nu}clei \underline{M}onitoring) project was proposed to carry 
out such monitoring observations for many AGNs 
(Kobayashi et al. 1998a; Yoshii 2002).  This project built a new 2-m 
optical-infrared telescope at the University of Hawaii's Haleakala 
Observatory on the Hawaiian island of Maui, and started preliminary 
observations in early 2001.  Light curve data of high photometric 
accuracy and frequent sampling have been accumulating since then. 

This paper reports early results for four nearby Seyfert 1 galaxies 
included in the MAGNUM program, including NGC 5548, NGC 4051, NGC 3227, 
and NGC 7469 for which the extent of the BLR and the central virial 
mass were previously derived from reverberation mapping of the BLR.  
The intensive monitoring observations using the MAGNUM telescope, 
with an improved technique for calculation of cross-correlation 
functions, enable the most accurate measurement of lag times between 
the $V$ and $K$ flux variations for the above four galaxies, and 
comparison with lag times for the broad emission lines. 

In \S \ref{obs} we describe the target AGNs, the observational 
conditions, the image reduction, and photometry procedures.  
In \S \ref{var} we discuss the characteristics of multicolor light 
curves of the target AGNs and their optical to near-infrared colors. 
We interpret the clear lag between optical and near-infrared 
variations in terms of dust reverberation. 
In \S \ref{ccfana} we perform a cross-correlation analysis to 
quantify the lag time between the $V$ and $K$ flux variations. 
Here, we describe how to calculate the cross-correlation functions 
based on two methods optimized for our time series data.
In \S \ref{discus} we discuss how well the measured lag times for 
the target AGNs relate to their optical nuclear luminosities and 
central masses, in combination with our earlier result of NGC 4151 
(Minezaki et al. 2004) and previous lag time measurements for other 
AGNs in the literature.  We compare these lag times with similar 
lag-time measurements for the broad emission lines, in order to see 
how the inner edge of the dust torus relates to the BLR.  
%Furthermore, we discuss the differences among successive lag time measurements and the luminosity dependence of the lag time for NGC 5548.

%%%%%%%%%%%%%%%%%%%%%%%%%%%%%%%%%%%%%%%%%%%%%%%%%%%%%%%%%%%%%%%%%%%%%%%%%%%%%%

\section{OBSERVATION AND DATA REDUCTION}
\label{obs}

%%%%%%%%%%%%%%%%%%%%%%%%%%%%%%%%%%%%%%%%%%%%%%%%%%%%%%%%%%%%%%%%%%%%%%%%%%%%%%

\subsection{Observation}

The characteristics of the target AGNs (NGC 5548, NGC 4051, NGC 3227, and 
NGC 7469) are listed in Table \ref{objlst}.  All of them are classified 
as Seyfert 1 galaxies. The right ascension and declination for an epoch 
of 2000 are given in columns 2 and 3, respectively. The apparent 
magnitude of average nuclear flux is given in column 4. The recession 
velocity and the Galactic $V$-band extinction are given in columns 
5 and 6, respectively, from the NED database\footnote{
The NASA/IPAC Extragalactic Database (NED) is operated by the Jet 
Propulsion Laboratory, California Institute of Technology, under 
contract with the National Aeronautics and Space Administration.
}. 
Assuming a Hubble constant of $H_0=70$ km s$^{-1}$Mpc$^{-1}$, the 
absolute $V$ magnitude estimated from column 4 is given in column 7, 
after being corrected for Galactic extinction in column 6, based on the average 
fluxes given in the next section (Table \ref{varstat}).  

The light curve data have been accumulated over the last several years, 
since the MAGNUM observatory started its operation in early 2001.  
A part of the early result has already been published as letters
(NGC 4151, Minezaki et al. 2004; NGC 5548, Suganuma et al. 2004).

Monitoring observations were made using the multi-color imaging 
photometer (MIP) mounted on the MAGNUM telescope (Kobayashi et al. 
1998a,b). The field of view of the MIP is 1.5 $\times$ 1.5 arcmin$^2$, 
and it enables us to obtain the images simultaneously in the optical 
and the near-infrared bands by splitting the incident beam into two 
different detectors such as an SITe CCD (1024 $\times$ 1024 pixels, 
0.277 arcsec/pixel) and an SBRC InSb array (256 $\times$ 256 pixels, 
0.346 arcsec/pixel). 

Monitoring observations in the $V$ and $K$ bands started in 2001, 
NGC 4051 in January, NGC 5548 in March, NGC 7469 in June, 
and NGC 3227 in November, while the $UBJH$ observations started 
later. The nuclear fluxes with the $V$ and $K$ filters were observed 
as the first priority, and the sampling interval for each target AGN 
was set shorter than one tenth of the expected lag time between the $V$ 
and $K$ flux variations.  The $B$ and $H$ filters were used as the 
second priority, and the $U$ and $J$ filters as the third priority.  

The telescope was pointed, typically six times, at each of 
the target AGN and two reference stars A and B, alternately.  
The pointings were repeated first in order of A $\rightarrow$ 
AGN $\rightarrow$ B, then in the reverse order with small 
shift of several arcsecs, and so on, which gave us dithered field
images of the AGN and the stars.
In each pointing, the CCD detector was exposed at once during which 
the InSb array was read out several times or more to avoid saturation.  
High signal-to-noise ratios were achieved with total on-source integration 
time of several minutes for each object.  

\begin{figure}
\epsscale{0.9}
\plotone{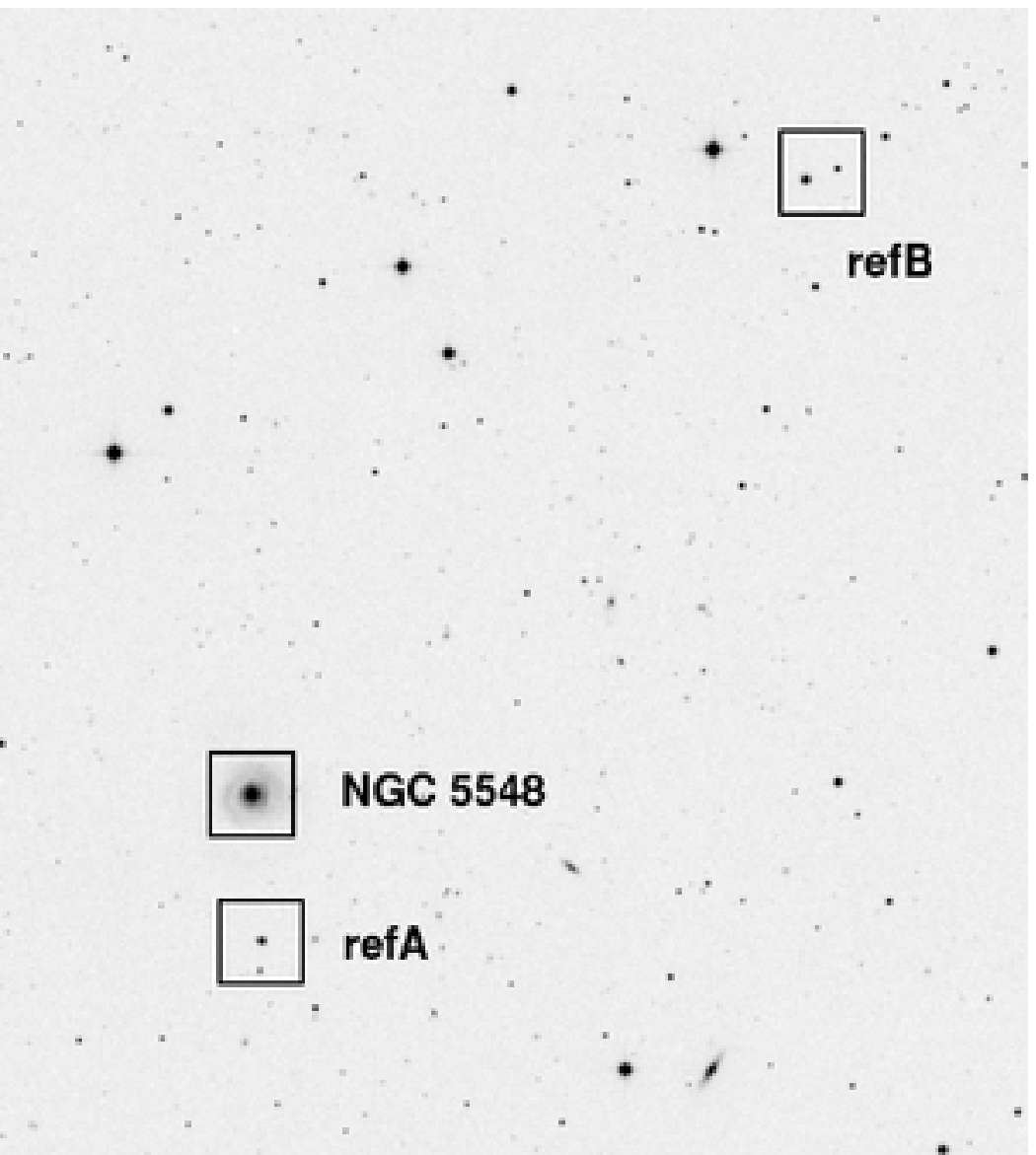}
\caption{Digitized Sky Survey (DSS) $R$-band image of the photometric 
     field for NGC 5548. The squares represent three $1'.5 \times 1'.5$ fields 
     of MIP camera. The left star (B1) in the ref B field is used for the $BVK$ 
     images, and the right star (B2) for the $H$ image.
\label{field_n5548}}
\end{figure}

\begin{figure}
\epsscale{0.8}
\plotone{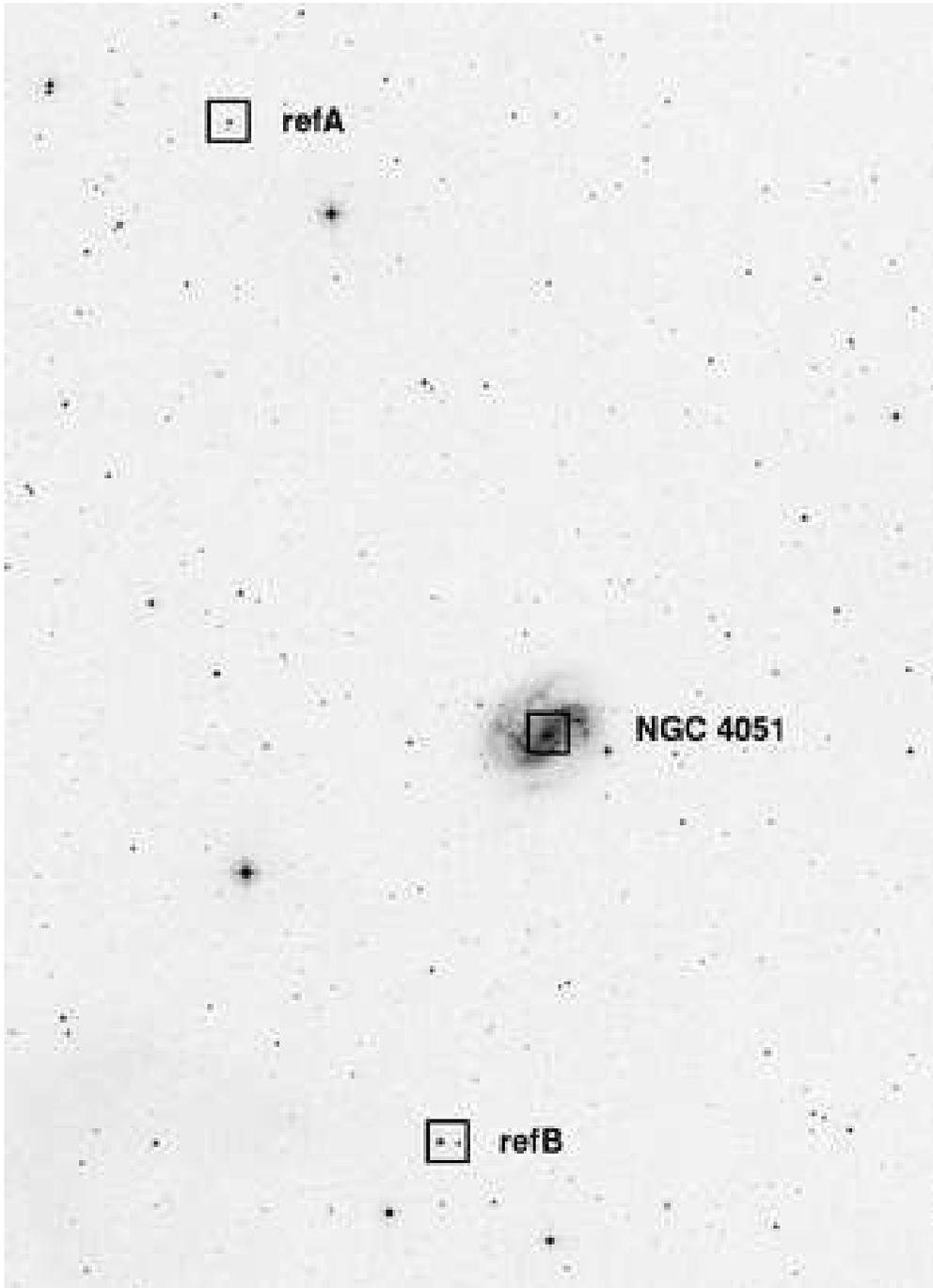}
\caption{DSS $B_j$-band image of the photometric field for NGC 4051. 
     The squares represent three $1'.5 \times 1'.5$ fields of MIP camera.
     The left star (B1) in the ref B field is used for the $JK$ images, 
     and the right star (B2) for the $BVH$ images. 
\label{field_n4051}}
\end{figure}

\begin{figure}
\plotone{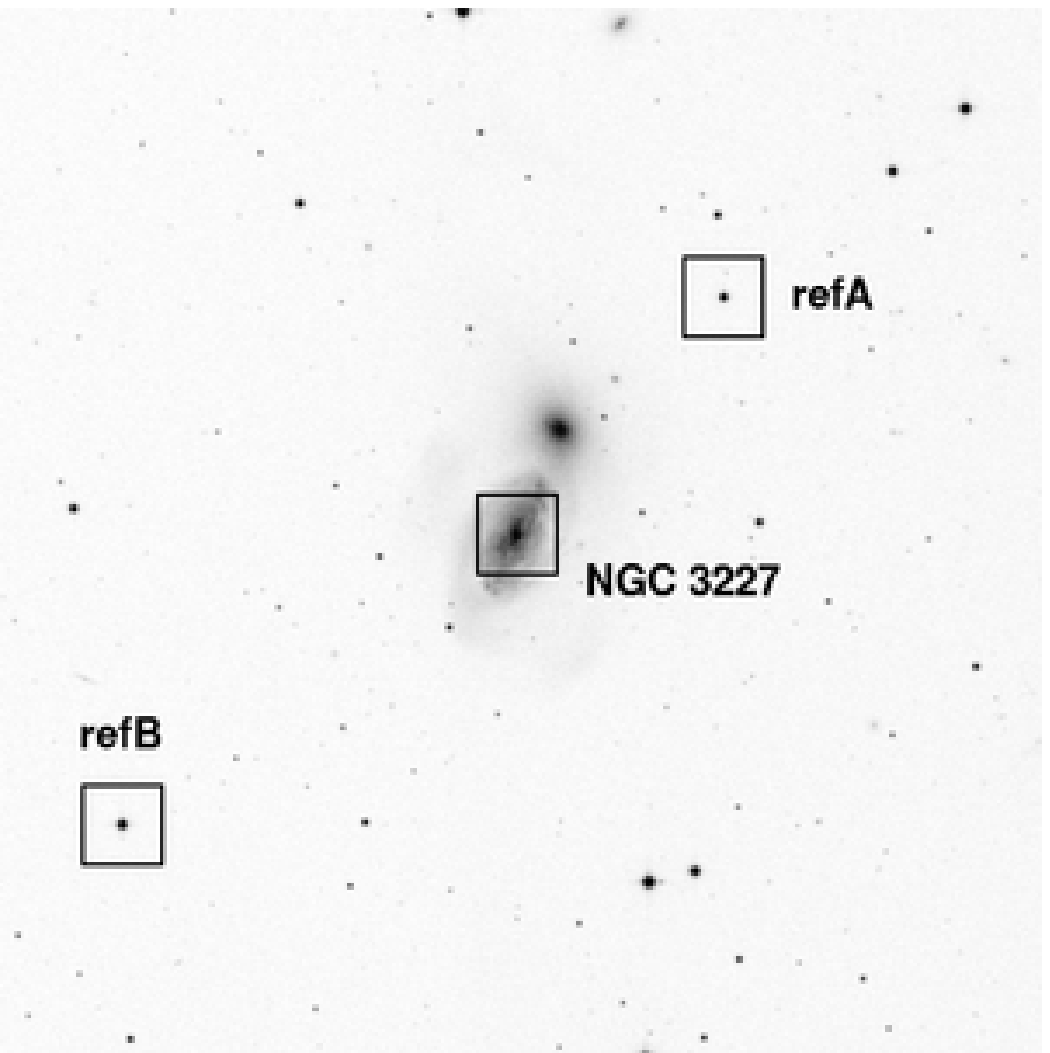}
\caption{DSS $B_j$-band image of the photometric field for NGC 3227. 
     The squares represent three $1'.5 \times 1'.5$ fields of MIP camera.
\label{field_n3227}}
\end{figure}

\begin{figure}
\epsscale{0.8}
\plotone{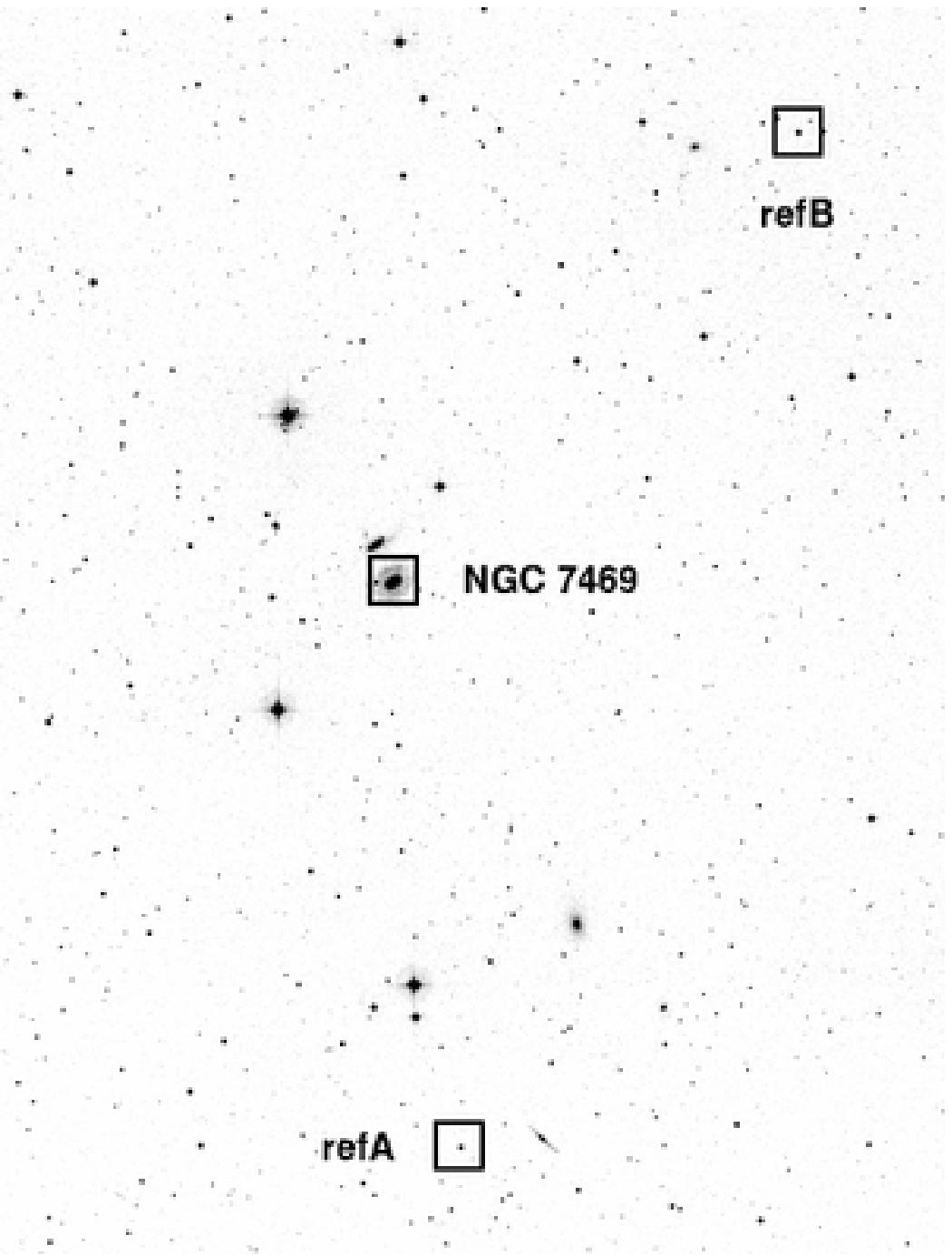}
\caption{DSS $R$-band image of the photometric field for NGC 7469. 
     The squares represent three $1'.5 \times 1'.5$ fields of MIP camera.
\label{field_n7469}}
\end{figure}

Digitized Sky Survey (DSS) images for four target fields are 
shown in Figures \ref{field_n5548} to \ref{field_n7469}.  
The FWHMs of the PSF during the observations were typically $1''.0-1''.5$ 
 in the optical and $0''.8-1''.0$ in the near-infrared.  
One standard star was observed each night to calibrate the flux of the 
reference stars. Twilight flat was obtained each evening for the $U$-band 
images, and dome flat for other bands was obtained at dawn after the observation.  
Our monitoring observations were performed with the highly automated observing 
system we developed (Kobayashi et al. 2003).

\subsection{Reduction and Photometry}

The images were reduced using IRAF\footnote{IRAF is distributed by 
the National Optical Astronomy Observatories, which are operated by 
the Association of Universities for Research in Astronomy, Inc., 
under cooperative agreement with the National Science Foundation.}.
We followed the standard procedures for image reduction, with small 
corrections for the non-linear response of the detectors. 
For the infrared image reduction, a sky image assembled from dithered 
images pointed at the reference stars was subtracted prior to 
flat-fielding. Then, flat-field corrections were applied using dome 
flats for the optical images and illumination-corrected dome flats 
for the near-infrared images. Twilight flats were used for the $U$-band 
images.

The nuclear fluxes of the galaxies were measured relative to the 
reference stars, and then the fluxes of these stars were calibrated. 
Aperture photometry within $\phi=8''.3$ with the sky reference 
of $\phi =11''.1-13''.9$ ring was applied to all images, then the 
fluxes of the nuclei and the reference stars were compared. We adopted 
a large aperture rather than that generally used for a PSF-like object 
because the variation of the PSF during the observation increases the 
scatter of the flux, which often dominates the uncertainties in the 
aperture photometry applied for the nucleus on the extended source of 
the host galaxy.

For each object and band, we used two comparison methods of 
relative photometry considering their conditions. In one method, the 
average fluxes of the nuclei and the reference stars over all dithering
sets were calculated before they were compared. In the other method, 
the fluxes of the nuclei and the reference stars were compared for each 
dithering set before the relative fluxes were assembled.  The latter method
can cancel relatively slow changes in transparency of the atmosphere 
during the observation, but has a statistical disadvantage because the 
number of the position is only six or less. We chose it mainly 
for the optical bands and chose the former mainly for the near-infrared bands.  
Rapid fluctuation of the flux between the dithering positions in the 
same night was attributed to thin cloud, and such observations were 
omitted from the light curve data. 

The uncertainty in the measurements is determined as the rms variations, 
divided by square root of the number of frames in the first method 
or the number of positions in the second method, about these averages 
in the relative photometry. The errors for each night were usually better 
than $\sigma =$ 0.01 mag. The flux ratios of two reference stars 
$f({\rm A})/f({\rm B})$ in various bands were monitored to confirm that 
they were not variable.

The fluxes of the reference stars were calibrated with respect to 
photometric standard stars taken from Landolt (1992) and Hunt et al. 
(1998). The magnitudes of the reference stars were determined using 
fixed airmass coefficients for many data sets for which the nucleus 
and the standard stars were observed in the same night. The airmass 
coefficients were stable on clear nights, and we adopted 
pre-measured values of 0.212 ($B$), 0.116 ($V$), 0.016 ($J$), and
0.011 ($H$) for unit airmass.  We adopted the fitted value for each 
calibration, ranging from 0.04 to 0.08 for the $K$ band.  
The calibration errors were less than 0.01 mag in most bands. 
The coordinates and magnitudes of the reference stars in the optical 
and near-infrared bands are shown in Tables \ref{refmagopt} and 
\ref{refmagir}, respectively.
The aperture magnitude of the nucleus was determined each night, relative 
to those of the reference stars obtained each night.  Small corrections 
for the color term in the optical filters were also applied.

The fluxes in the aperture are significantly contaminated by starlight 
of the host galaxy.  The offset galaxy fluxes for the target AGNs 
were estimated as follows: The surface profile of the galaxy was 
fitted using GALFIT (Peng et al. 2002) where the images of the 
reference stars in the same night were used as the PSF.  This profile 
was subtracted from the observed image, and the flux of the residual 
PSF-like image was measured.  Then, the offset galaxy flux in the 
aperture was determined as the difference between the total flux in 
the aperture and the flux of the residual PSF-like image. 
The offset galaxy flux estimated from the data of best PSF was 
subtracted from the observed light curve.

In the case of NGC 7469, which has a ring-shaped star-forming region with a
radius of about $1''.3$ around the nucleus, the flux of the PSF-like image was 
estimated without obtaining the galaxy profile with GALFIT.
Instead, we estimated the nuclear flux by substracting the 
central PSF image from the observed image and regarding the
residual at the center as the galaxy offset.  A lower nuclear 
flux was then obtained from a larger offset by equating such 
residual with the brightness at the radius of the star-forming 
ring, while a larger nuclear flux from a lower offset by equating 
such residual with the brightness at the radius outside which the 
galaxy profile dominates over the profile of star-forming ring.

The offset galaxy fluxes thus estimated in the $\phi=8''.3$ are given  
in Table \ref{galfx}. Their uncertainty estimated from the scatter 
over different nights or the residual flux at the origin of the flux to 
flux diagram is $10-30$ \%, sometimes giving a significant error in 
the nuclear flux which may result in false variations of the nuclear color.  
Note that the galaxy subtraction provides an offset flux only and does 
not affect the cross-correlation analysis in \S \ref{ccfana}.

\begin{figure}
\plotone{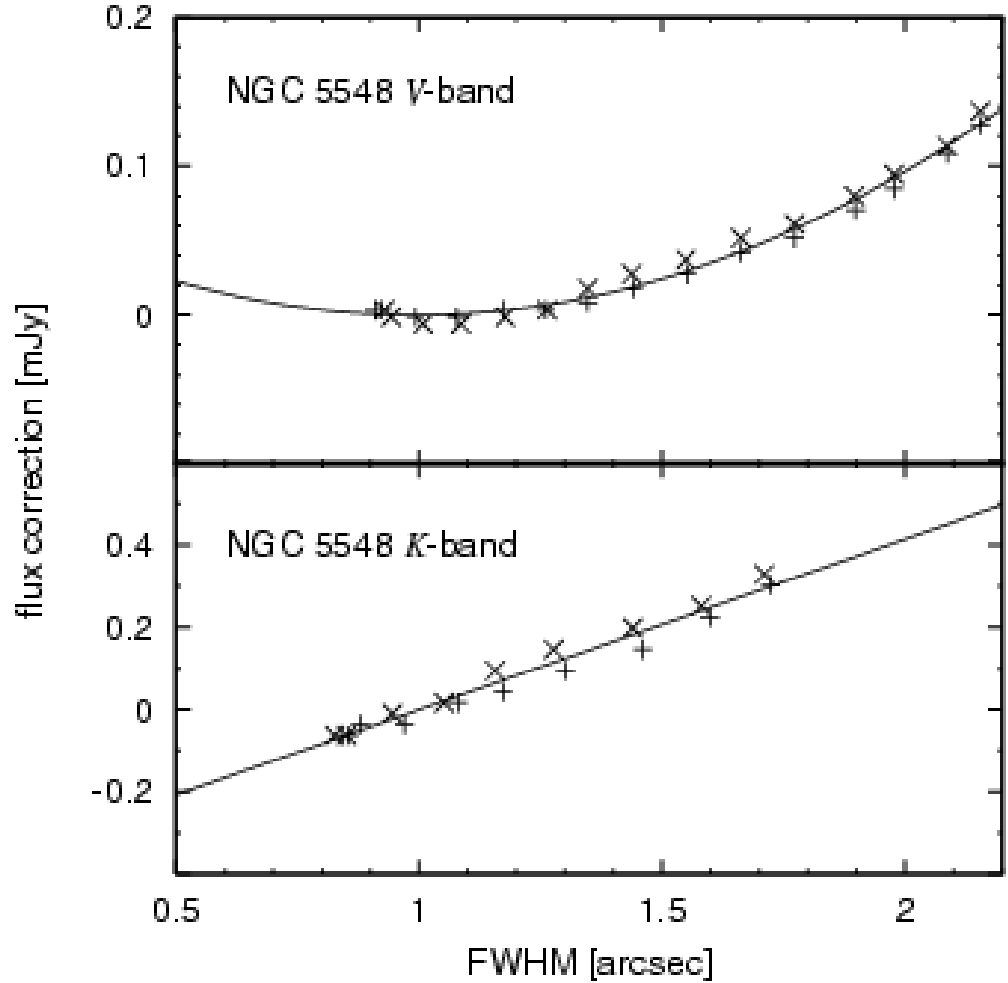}
\caption{Correction of nuclear flux for the seeing effect, scaled to 
     be zero at ${\rm FWHM}=1''$, for some example cases of NGC 5548.  
     Seeing-dependence of nuclear flux is simulated by convolving 
     the image of good seeing conditions with the FWHM chosen.  Different 
     symbols show the simulated results for different epochs. Thick 
     lines are the fitting curves.
\label{seef_gal}}
\end{figure}

The host galaxy flux in the aperture depends on the seeing FWHM. 
This effect makes a systematic change in the measured nuclear flux
up to 0.03 mag depending on the central profile of the host galaxy,
and causes false fluctuations in the light curve.  
In order to estimate a relation between the seeing FWHM and the 
corresponding nuclear flux, the images in the best seeing condition 
were convolved with several FWHM values, as shown in Figure 
\ref{seef_gal}.  Thus, the nuclear flux measured each night was 
corrected using the fitting formulae:   
\begin{equation}
\Delta F = a({\rm FWHM}-1) + b({\rm FWHM}-1)^2 \;\;[{\rm mJy}],
\label{seecor_eq}
\end{equation}
where $a$ and $b$ are the constants and their values for each nucleus
and band are given in Table \ref{seecor}.  However, such correction 
was not attempted in the cases of $BJH$ for NGC 4051 and $BVJHK$ for 
NGC 3227, because the photometric errors of each night were found to 
dominate over the seeing effect.

\begin{figure}
\epsscale{1.05}
\plotone{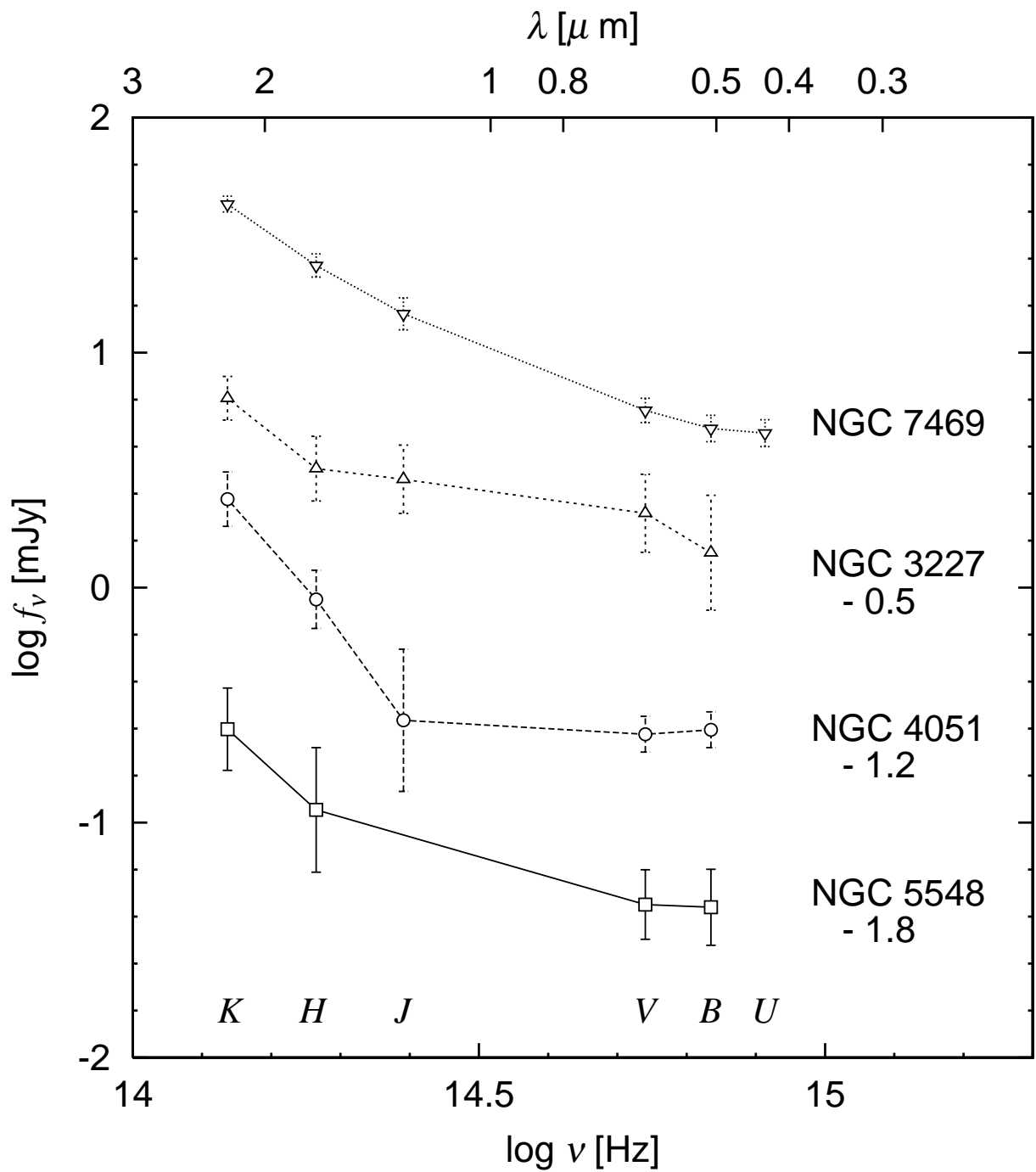}
\caption{Optical to near-infrared spectral energy distributions (SEDs) 
    for the target nuclei, where the host galaxy contribution has been 
    subtracted.  The fluxes shown are averaged over all observing 
    nights.  The error bars include the rms flux variations and 
    subtraction errors.  The number below each object name is the 
    offset value to log$f_{\nu}$.
\label{sed}}
\end{figure}

Figure \ref{sed} shows the spectral energy distribution (SED) observed 
in the optical ($UBV$) and near-infrared ($JHK$) for each nucleus.
The fluxes are those averaged over the nights when the SEDs were 
observed.  The error bars include their rms variations and the 
uncertainties in the offset subtraction.  We see that the 
SED is steep in the near-infrared ($JHK$) while it is flat in the 
optical ($UBV$) except for NGC 3227.  This is consistent with a view
that the hot dust emission dominates towards longer wavelengths in the 
near-infrared, while the flat power-law spectra from nuclei of Seyfert 
galaxies dominates in the UV and optical.  However, NGC 3227 shows a
steep SED in the optical which is different from the other nuclei.
This may originate from offset errors in the $V$ band (see \S \ref{var_opt}),
rather than originating from the central engine .

\begin{figure}
\epsscale{0.95}
\plotone{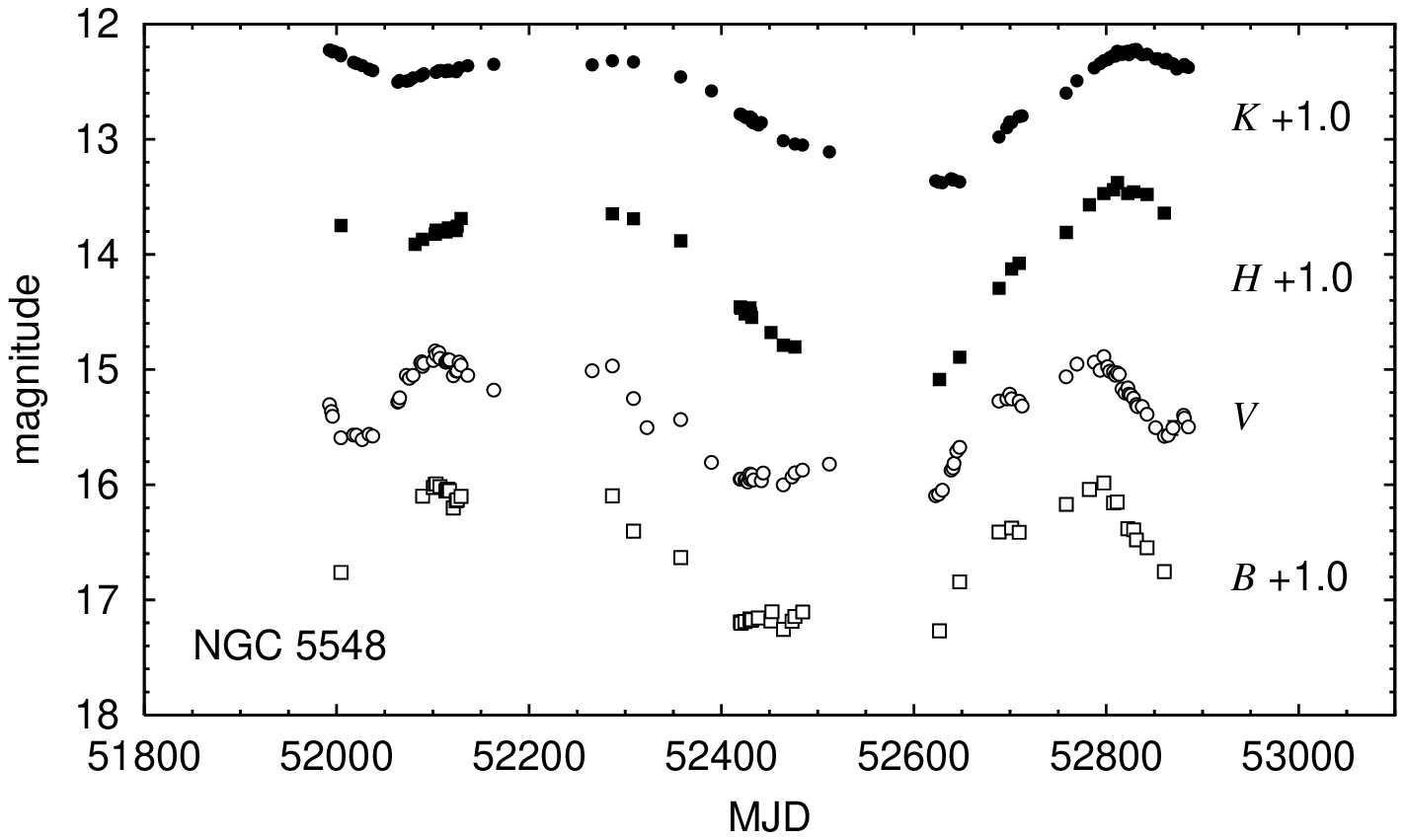}
\caption{Observed $BVHK$ light curves of NGC 5548 nucleus for the period 
    from March 2001 to September 2003. 
\label{lc_ngc5548}}

\clearpage

\epsscale{0.95}
\plotone{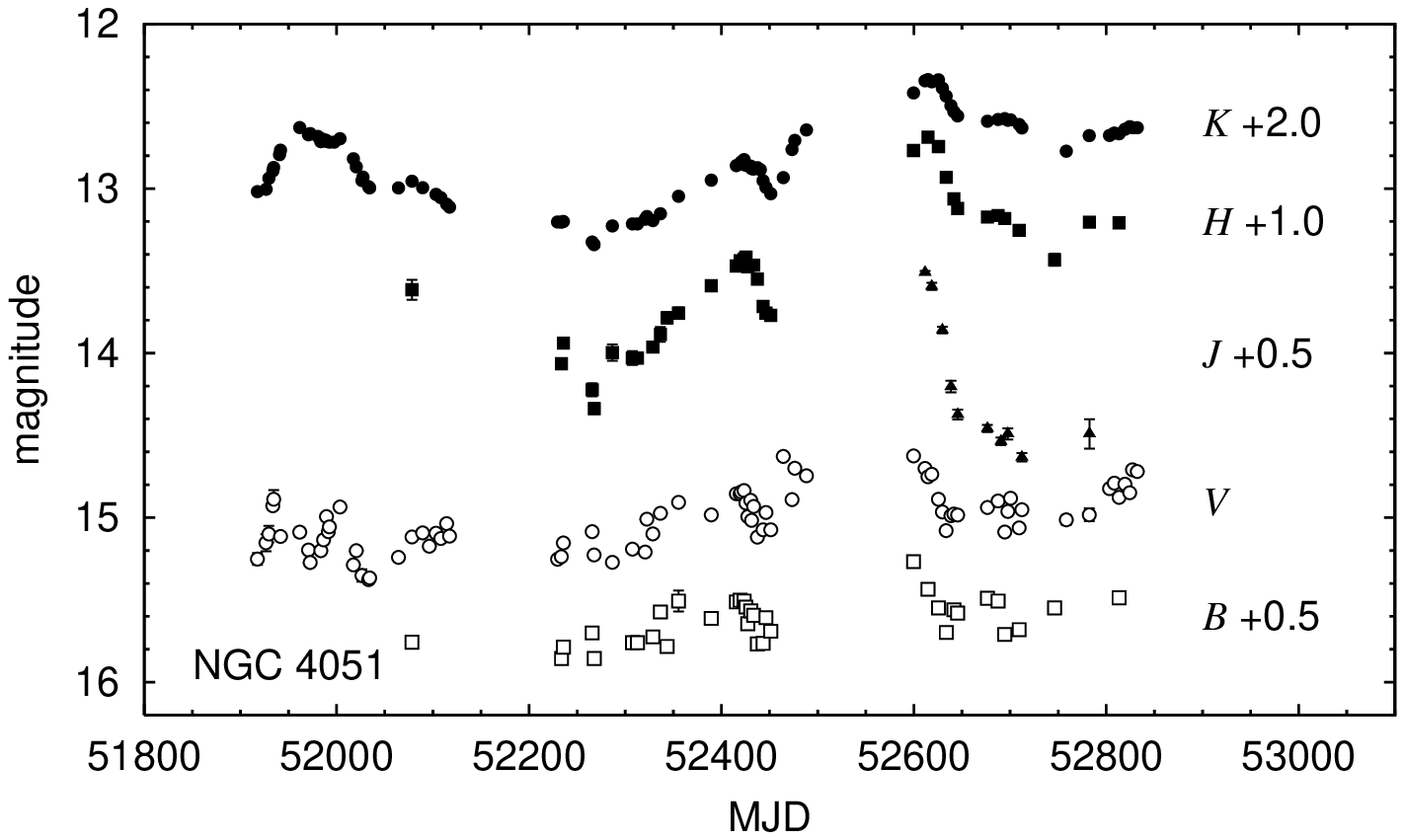}
\caption{Observed $BVJHK$ light curves of NGC 4051 nucleus for the period
    from January 2001 to July 2003. 
\label{lc_ngc4051}}
\end{figure}

\begin{figure}
\epsscale{0.95}
\plotone{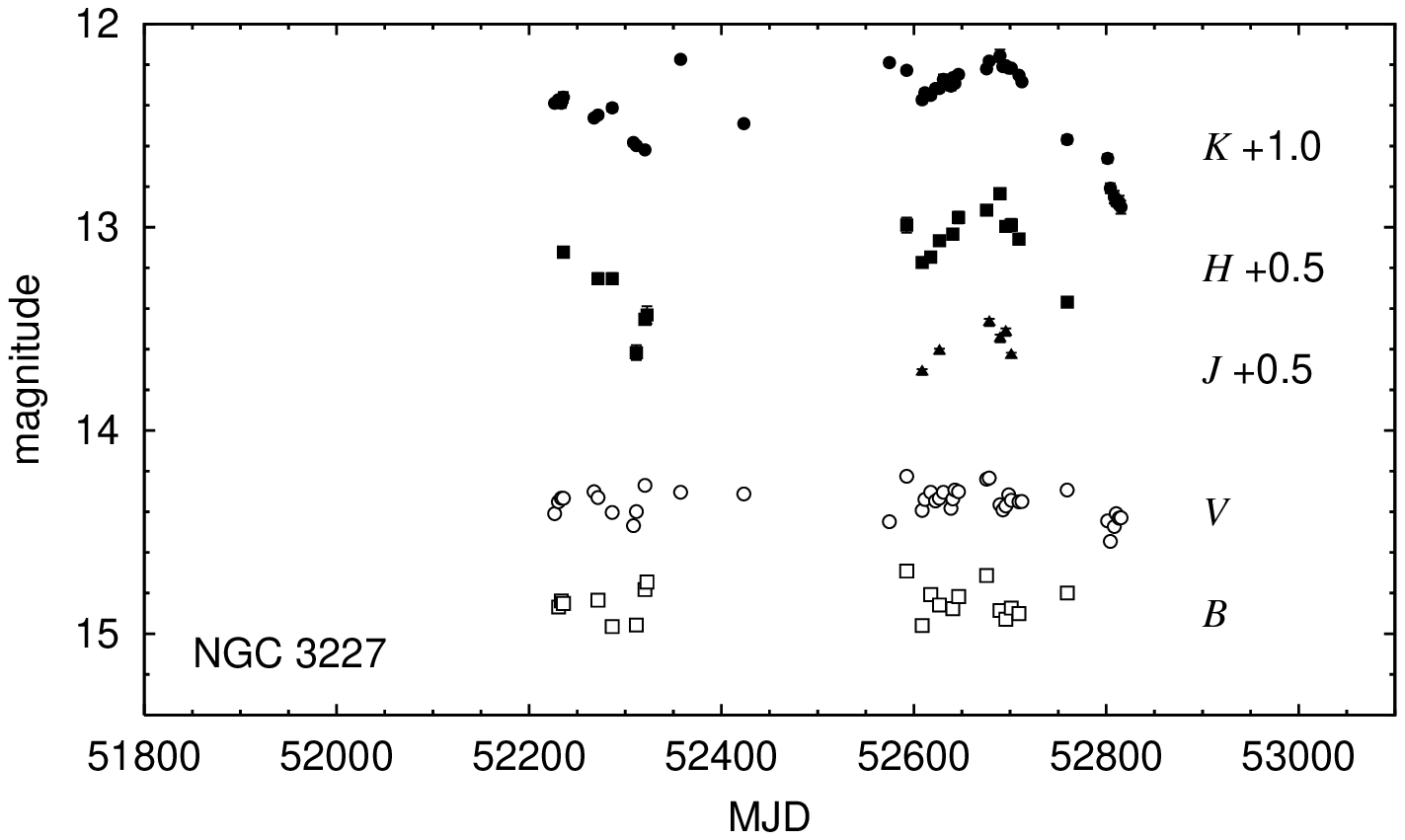}
\caption{Observed $BVJHK$ light curves of NGC 3227 nucleus for the period
    from November 2001 to June 2003. 
\label{lc_ngc3227}}

\clearpage

\epsscale{0.95}
\plotone{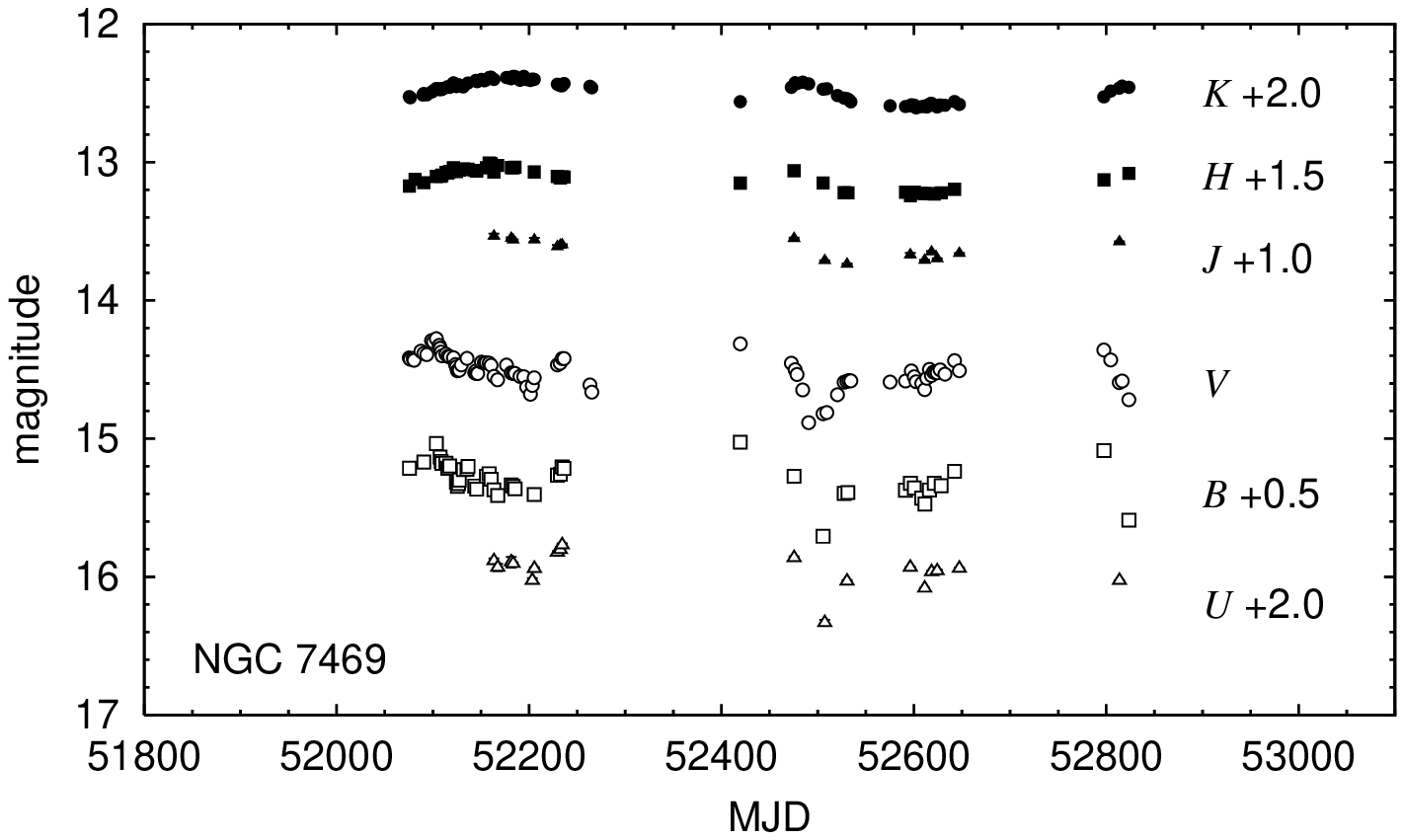}
\caption{Observed $UBVJHK$ light curves of NGC 7469 nucleus for the period
    from June 2001 to July 2003. 
\label{lc_ngc7469}}
\end{figure}

The light curves based on aperture fluxes of four target AGNs are shown in 
Figures \ref{lc_ngc5548} to \ref{lc_ngc7469}. The correction for the seeing
effect has been applied and the offset flux of the host galaxy has been 
subtracted here.

%%%%%%%%%%%%%%%%%%%%%%%%%%%%%%%%%%%%%%%%%%%%%%%%%%%%%%%%%%%%%%%%%%%%%%

\section{Features of Optical and Near-infrared Variations}
\label{var}

%%%%%%%%%%%%%%%%%%%%%%%%%%%%%%%%%%%%%%%%%%%%%%%%%%%%%%%%%%%%%%%%%%%%%%

\subsection{Data}

The sampling characteristics of the $V$ and $K$ light-curve data are given in 
Table \ref{smpstat}. 
The whole monitoring period for each galaxy and the maximum intervals 
between adjacent sampling epochs are given in columns 2 and 3, respectively. 
The number of data points, the average and median intervals for the $V$
band are given in columns 4, 5 and 6, respectively, and those for the $K$
band are in columns 7, 8 and 9.
The maximum interval corresponds to the sampling gap due to solar conjunction. 
The average interval is longer than the median, because of sampling 
gaps of three months due to solar conjunction and occasional blanks of 
several weeks caused by bad weather or facility maintenance. 
The median interval of $3-5$ days over $2-3$ years is much shorter than any 
other monitoring observations in the near-infrared.  Because of changing 
weather conditions and/or sudden problems with instruments, there are 
cases where the observations in the $V$ band were successful while not 
in the $K$ band, and vice versa. Therefore, the number of $V$-band 
data points does not necessarily coincide with that for the $K$ band. 

The statistics that characterize the observed variability in the $V$ 
and $K$ bands are given in Table \ref{varstat}.  The values in this 
table were calculated after the offset flux of the host galaxy in 
the aperture was subtracted.  For each target AGN, the average 
flux $\left<f\right>$, the maximum to minimum flux ratio $R_{\rm max}$,
and the normalized variability amplitude $F_{\rm var}$ are given.
Here, we define $F_{\rm var}$ as the variance of the flux $\sigma$, 
corrected for the measurement uncertainty $\delta$, divided by the
average flux: 
\begin{equation}
F_{\rm var}=\frac{\sqrt{\sigma^2-\delta^2}}{\left<f\right>} \;\;\;,
\label{mf_eq}
\end{equation}
where
\begin{equation}
\sigma^2=\frac{1}{N}\sum_{i=1}^{N}(f_{i}-\left<f\right> )^2 \;\;\;,
\end{equation}
and
\begin{equation}
\delta^2=\frac{1}{N}\sum_{i=1}^{N}\delta_{i}^2 \;\;\;.
\end{equation}
It is evident from the values of $F_{\rm var}$ in this table that 
the four AGNs underwent significant $V$ and $K$ variations during our 
monitoring observations.

\subsection{Optical Variation}
\label{var_opt}

\begin{figure}
\epsscale{1.0}
\plotone{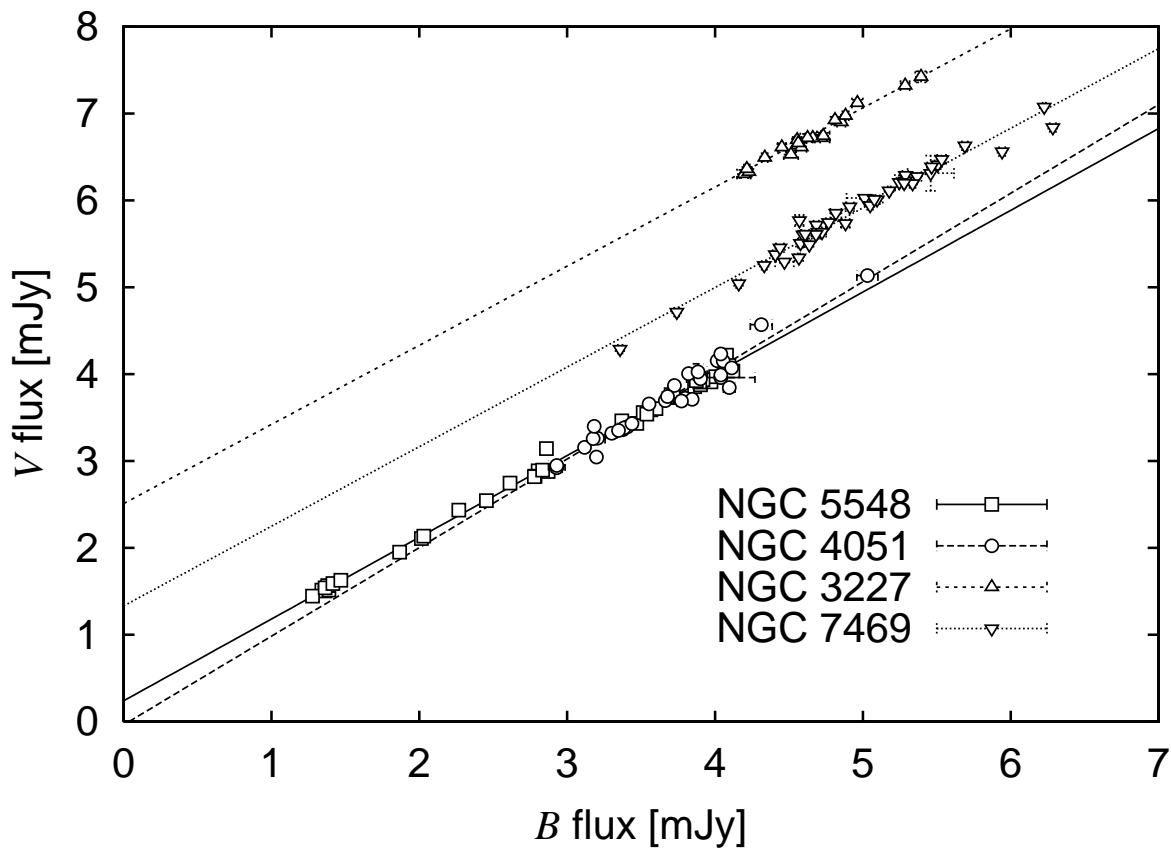}
\caption{The flux-to-flux diagram where the $V$ flux is plotted 
    against the $B$ flux for the target AGNs. The host galaxy flux, 
    given in Table \ref{galfx}, has been subtracted from the nuclear 
    flux.  The error bars correspond to the errors of differential 
    photometry in deriving the light curves.  Lines represent linear 
    regressions.
\label{fvg_BV}}
\end{figure}

Comparison of flux variations between the $B$ and $V$ bands for 
each AGN indicates that their long timescale variations as well as small, 
short timescale fluctuations are almost synchronized (Figures 
\ref{lc_ngc5548} to \ref{lc_ngc7469}).  This tendency is also 
true of the $U$-band variations for NGC 7469.  
Figure \ref{fvg_BV} shows the $V$ flux plotted against the $B$ 
flux for each AGN.  Each point represents a pair of nuclear 
fluxes in a certain night from which the flux of the host galaxy 
has been subtracted, and the error bars represent the photometric
errors.  There is a tight linear relation between the flux 
variations in the $B$ and $V$ bands. 

Winkler et al. (1992) and Winkler (1997) reported this  
linear relation between the optical flux variations for many 
Seyfert galaxies in the southern hemisphere. They considered that 
the slope of the linear relation in the flux-to-flux diagram, 
called a flux variation gradient (FVG), would represent the actual 
color of the variable component, which appears to be constant 
for individual AGNs.  If this is the case, the offset of the 
linear regression line from the origin is due to non-variable 
components, such as errors in subtracting the galaxy flux, 
unresolved narrow-line fluxes, etc. 

A tight linear relation and no sign of significant non-linearity 
in Figure \ref{fvg_BV} support the above view, and the color of the 
central engine derived from the FVG, which is unaffected by 
non-variable components, would be more reliable than those inferred 
from the SED in Figure \ref{sed}. Our estimate of nuclear $B-V$ color from the 
FVG, after being corrected for Galactic extinction, gives 
$(B-V)_{\rm FVG}=0.04\pm0.01$ (NGC 5548), $0.13\pm0.05$ 
(NGC 4051), $0.00\pm0.04$ (NGC 3227), and $-0.04\pm0.03$ (NGC 7469). 
These values are similar to a typical value for Seyfert 1 
galaxies reported by Winkler et al. (1992), and correspond to 
the power-law nuclear flux $f_{\nu}\propto\nu^{\alpha}$ with 
$\alpha=+0.36\pm0.03$ (NGC 5548), $-0.04\pm0.22$ (NGC 4051), 
$+0.53\pm0.17$ (NGC 3227), and $+0.70\pm0.11$ (NGC 7469), 
provided the effects of emission-line fluxes such as H$\beta$
are not taken into account.

\subsection{Near-infrared Variation}
\label{var_ir}

In contrast with the optical light curves showing the short-term 
fluctuations, the near-infrared light curves in the $JHK$ bands 
are smooth (Figures \ref{lc_ngc5548} to \ref{lc_ngc7469}).  
This is consistent with a view that the bulk of near-infrared 
emission comes from the thermal reprocessing region whose size 
exceeds the distance over which light travels in the timescale 
of short-term fluctuations in the optical.

\begin{figure}
\epsscale{1.0}
\plotone{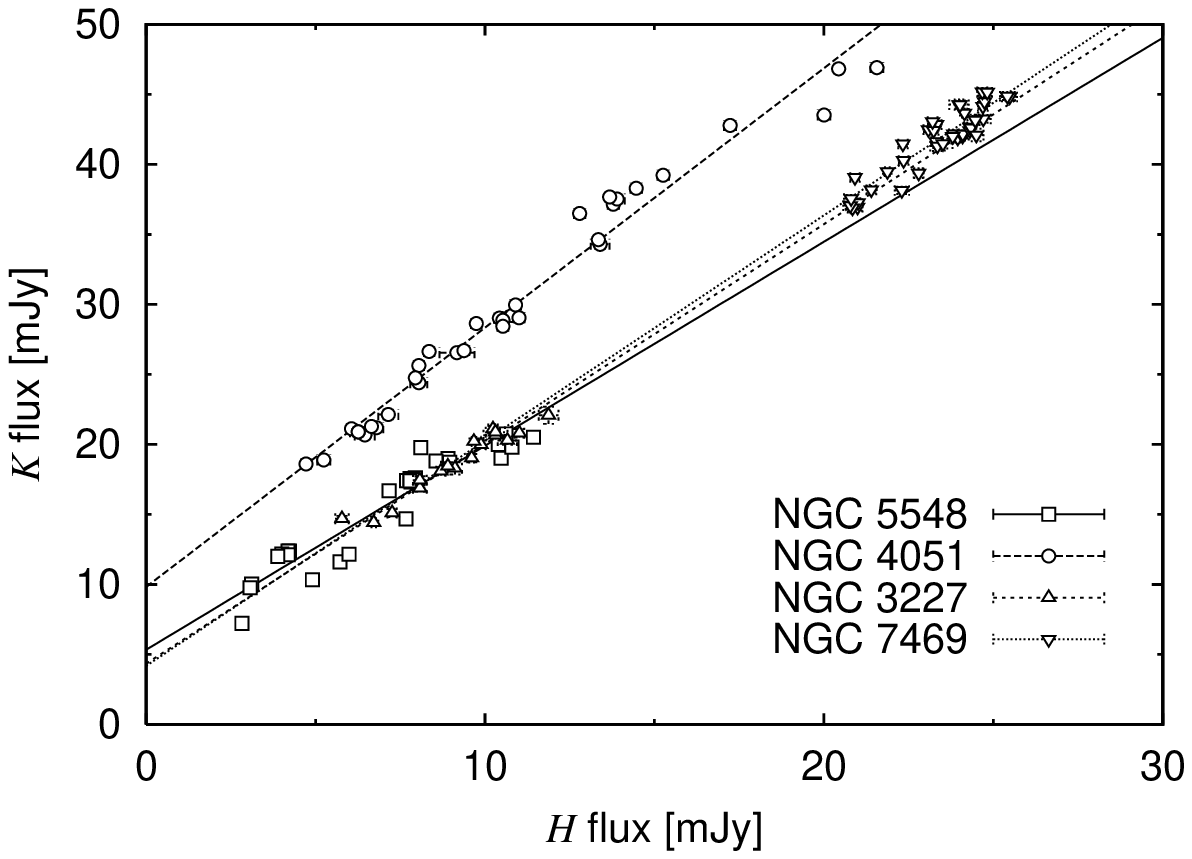}
\caption{The flux-to-flux diagram.  The same as Figure 
    \ref{fvg_BV}, except that the $K$ flux is plotted against the 
    $H$ flux for the target AGNs.  
\label{fvg_HK}}
\end{figure}

Figure \ref{fvg_HK} shows the $K$ flux plotted against the $H$ flux 
for each AGN.  There is a linear relation between the flux variations 
in the $K$ and $H$ bands.  Our estimate of nuclear $H-K$ color from 
the FVG gives $(H-K)_{\rm FVG}=0.92\pm0.07$ (NGC 5548), $1.17\pm0.04$ 
(NGC 4051), $1.00\pm0.06$ (NGC 3227), and $1.02\pm0.07$ (NGC 7469), 
which is consistent with the FVG colors for dozens of Seyfert galaxies measured 
by Glass (2004).
These colors correspond to the black-body temperature of $1500-1800$ K which 
agrees with the sublimation temperature of graphite grains, and 
support the thermal origin of near-infrared emission, especially in the 
$K$ band.

There is a scatter in the FVG in Figure \ref{fvg_HK}, and the relation 
between the $H$ and $K$ flux variations is not as tight as that for
the $B$ and $V$ fluxes. This implies that the thermal reprocessing 
region that emits the $H$ flux may not completely coincide with that
for the $K$ flux, and/or the central source could contaminate differently
in the $H$ and $K$ fluxes.

\subsection{Correlation and Lag time between Optical and Near-infrared 
Variations}
\label{OI}

\begin{figure}
\epsscale{1.0}
\plotone{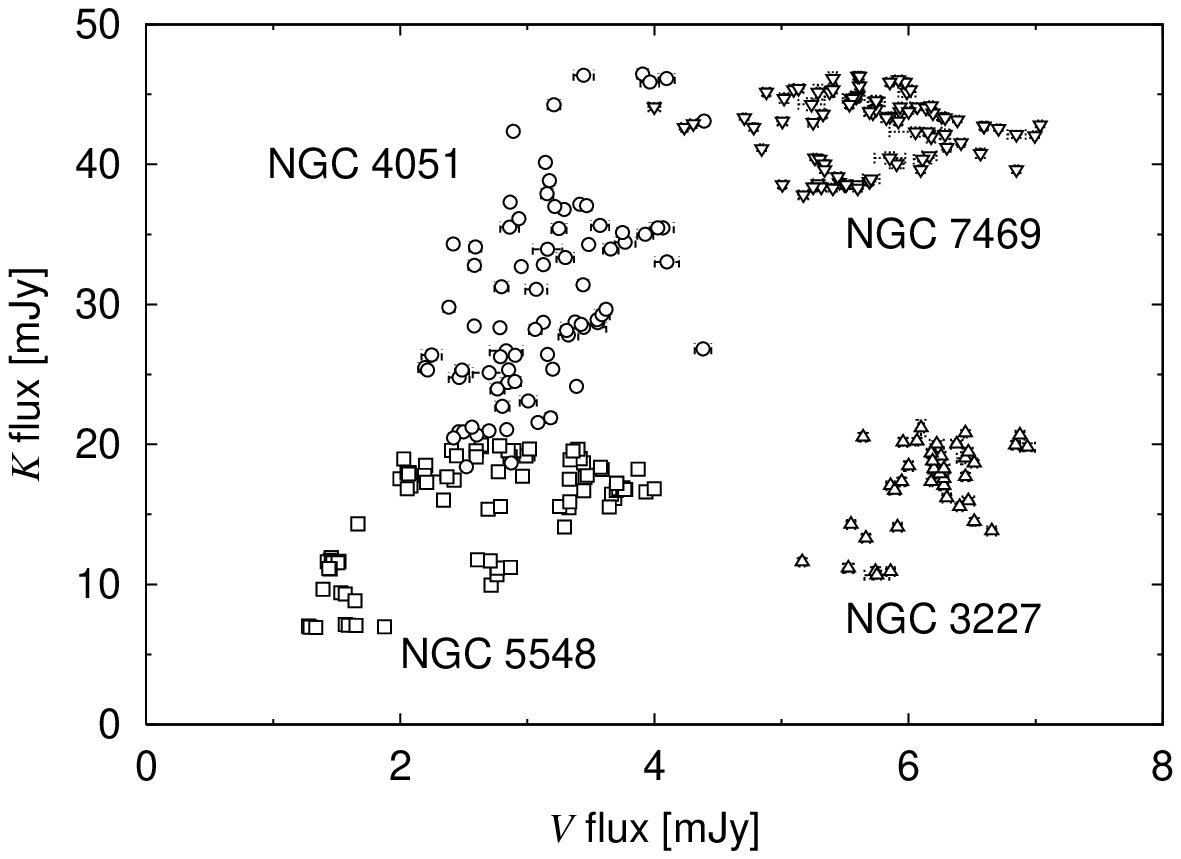}
\caption{The flux-to-flux diagram.  The same as Figure 
    \ref{fvg_BV}, except that the $K$ flux is plotted against the $V$ flux 
    for the target AGNs.
\label{fvg_VK}}
\end{figure}

The most important feature of the near-infrared variation, besides 
its smoothness, is its lag time behind the optical variation. 
Figures \ref{lc_ngc5548} to \ref{lc_ngc7469} show that the 
near-infrared light curve resembles the optical light curve, 
if shifted backwards by about ten days (NGC 4051 and NGC 3227) or 
several tens of days (NGC 5548 and NGC 7469).  Figure \ref{fvg_VK} 
shows the $K$ flux plotted against the $V$ flux for each AGN. 
Because of the short-term fluctuations present only in the $V$ band,
and because of the delay between the $V$ and $K$ variations, there 
is a significant scatter in Figure \ref{fvg_VK}. 

\begin{figure}
\epsscale{0.95}
\plotone{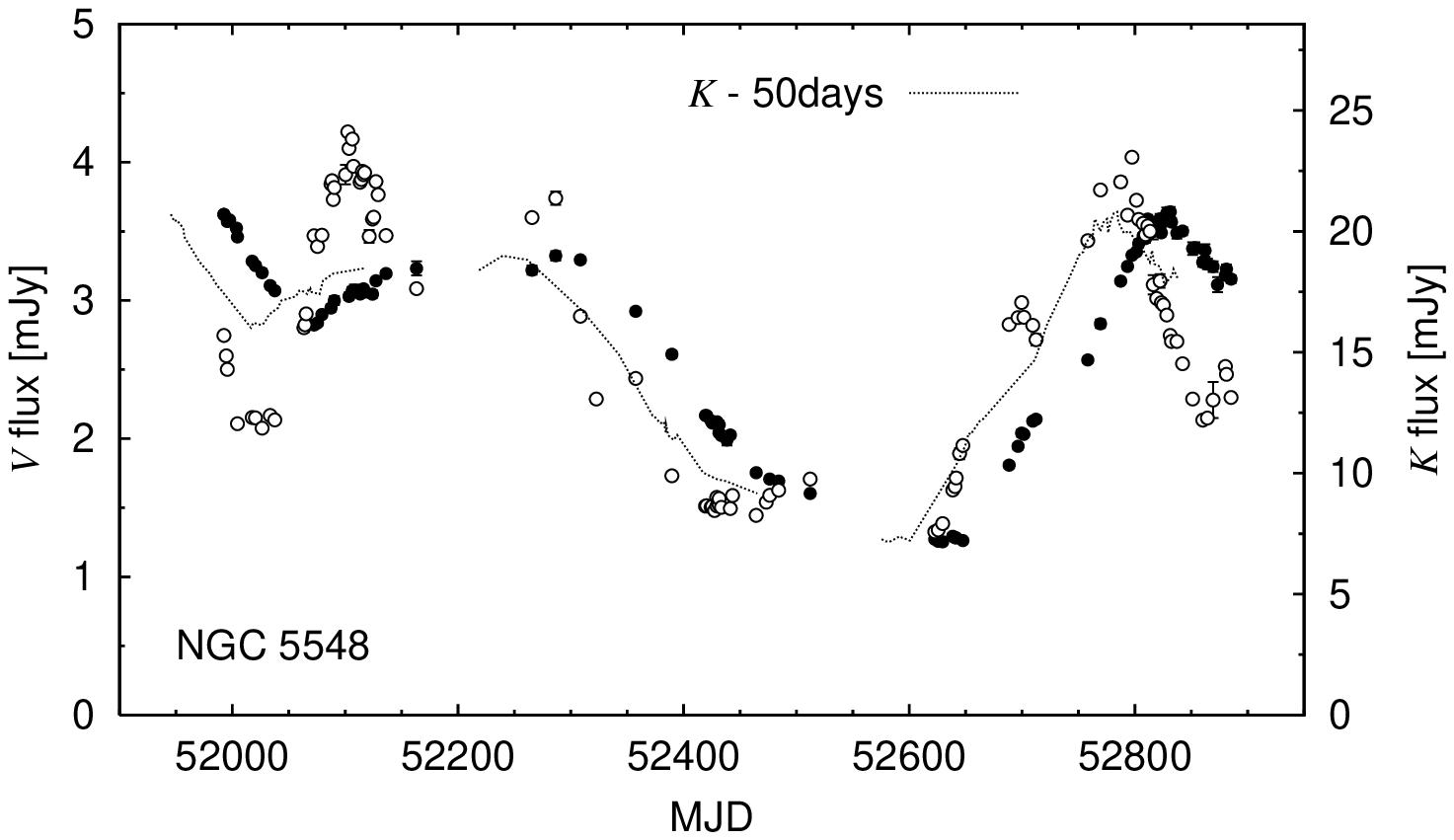}
\caption{Observed $V$ and $K$ light curves of NGC 5548 nucleus.  
    Open and filled circles correspond to $V$ and $K$, 
    respectively.  The dotted line is the $K$ light curve 
    shifted backwards by 50 days.
\label{lc_ngc5548_VK}}

\clearpage

\epsscale{0.95}
\plotone{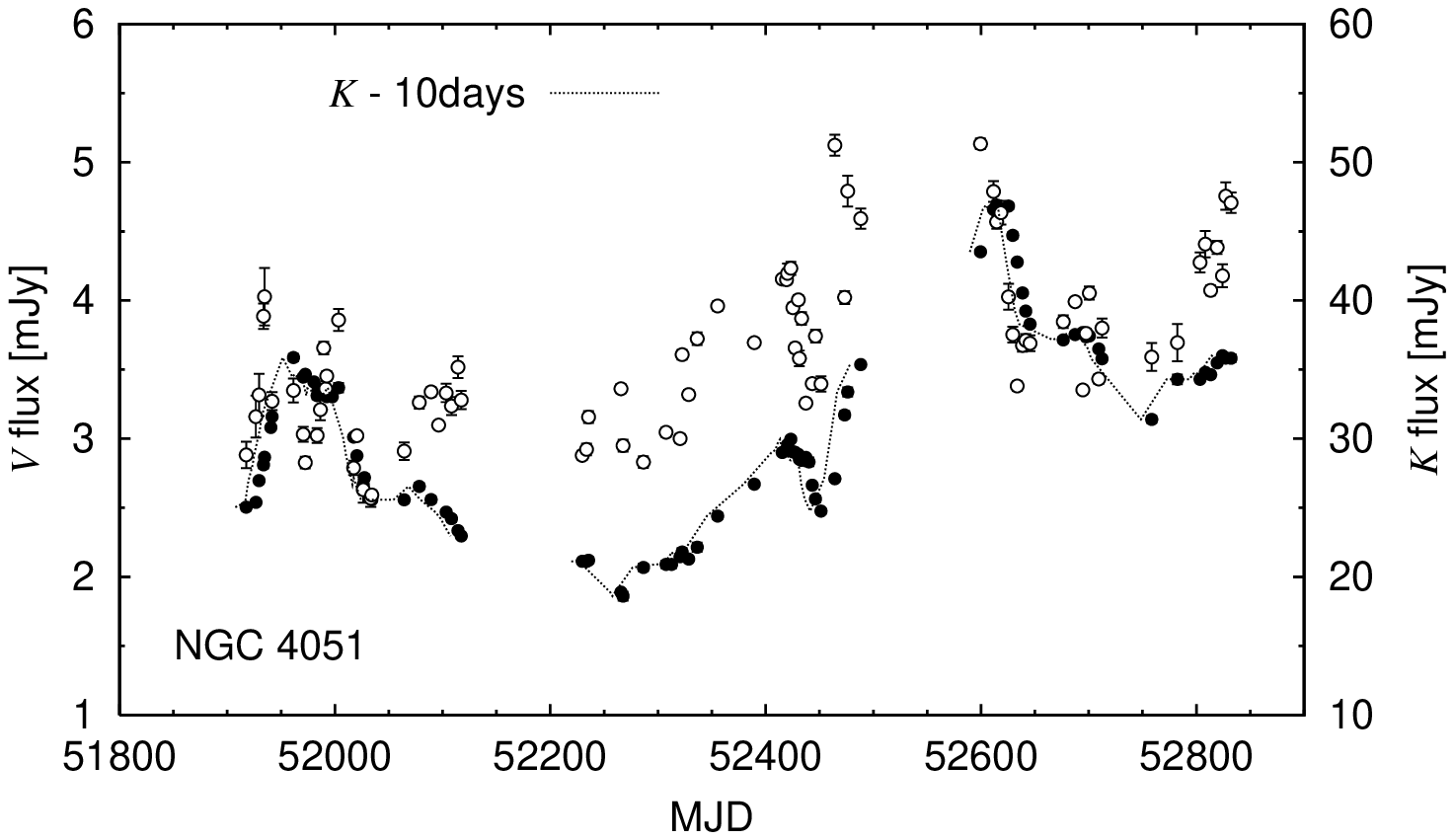}
\caption{Observed $V$ and $K$ light curves for NGC 4051 nucleus.
    Others are the same as Figure \ref{lc_ngc5548_VK}, but for the 
    shift by 10 days.
\label{lc_ngc4051_VK}}
\end{figure}

\begin{figure}
\epsscale{0.95}
\plotone{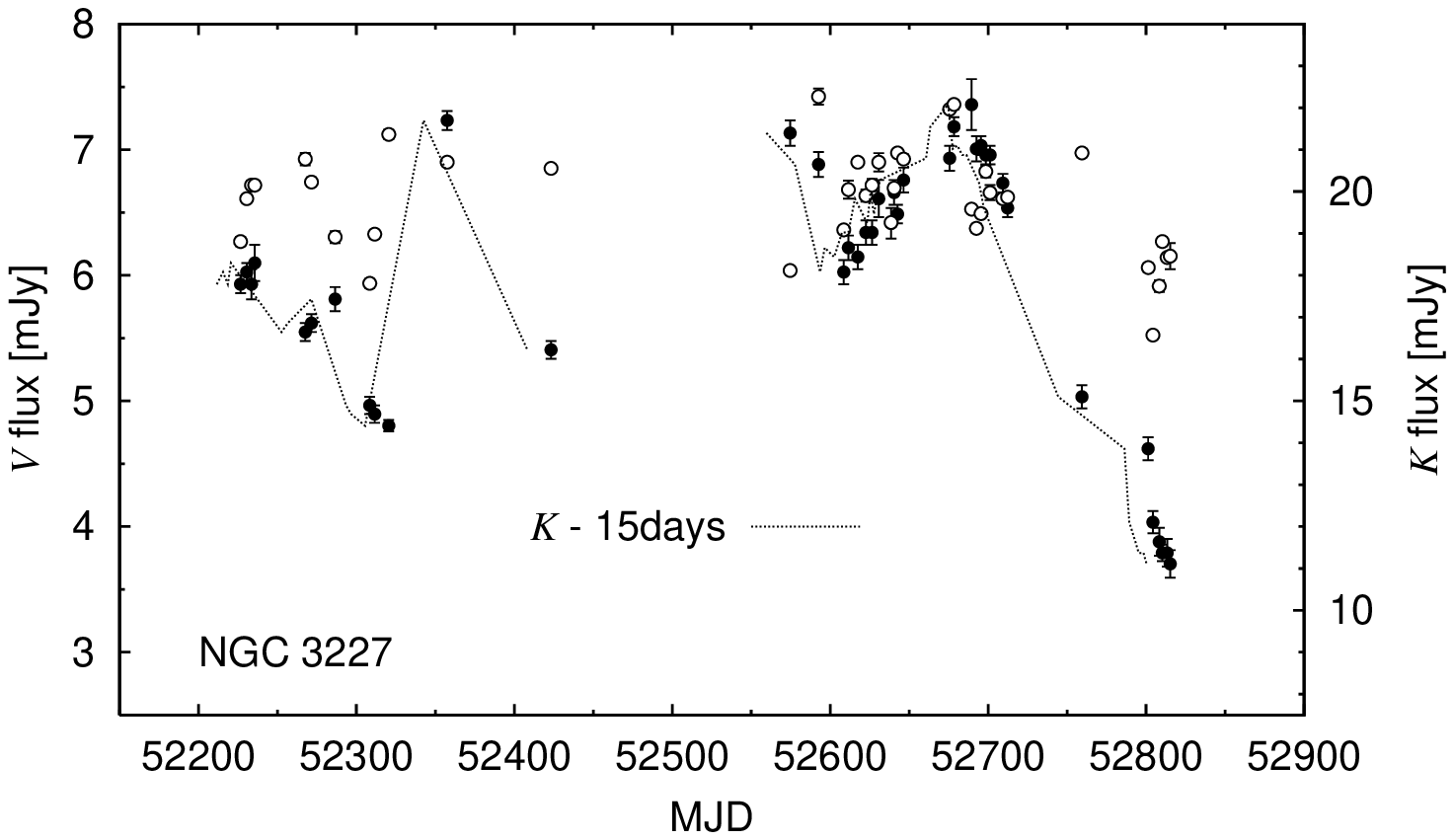}
\caption{Observed $V$ and $K$ light curves for NGC 3227 nucleus.
    Others are the same as Figure \ref{lc_ngc5548_VK}, but for the 
    shift by 15 days.
\label{lc_ngc3227_VK}}

\clearpage

\epsscale{0.95}
\plotone{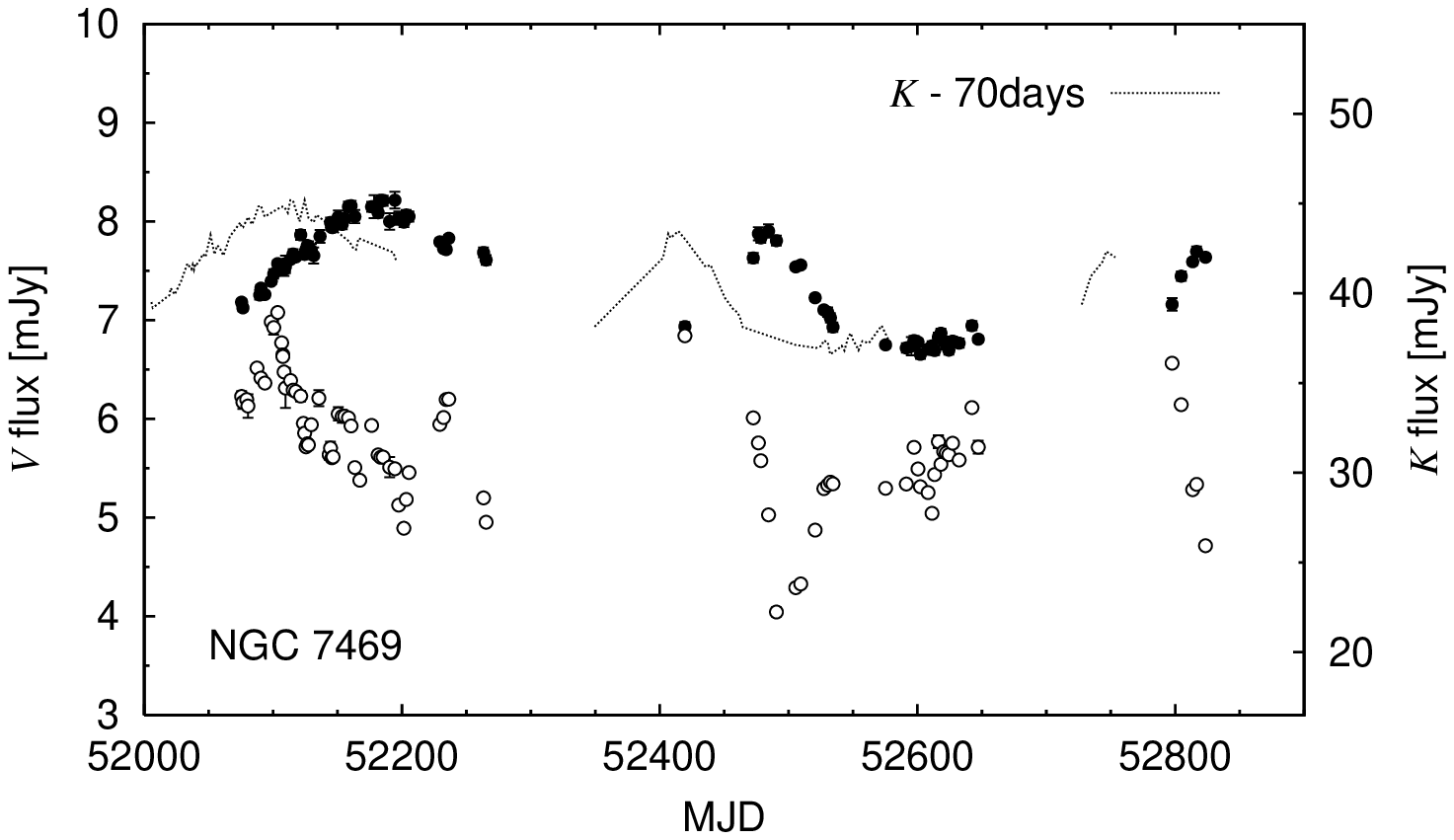}
\caption{Observed $V$ and $K$ light curves for NGC 7469 nucleus.
    Others are the same as Figure \ref{lc_ngc5548_VK}, but for the 
    shift by 70 days.
\label{lc_ngc7469_VK}}
\end{figure}

The lag time is approximately estimated by shifting either one of the 
$V$ or $K$ light curves horizontally.  Figures \ref{lc_ngc5548_VK} to
\ref{lc_ngc7469_VK} show the observed $V$ and $K$ light curves for each 
AGN.  The dotted curve in each figure is the $K$ light curve shifted 
backwards by the lag time determined approximately by sight. A more 
reliable estimate of lag time will be given in the next section.

%%%%%%%%%%%%%%%%%%%%%%%%%%%%%%%%%%%%%%%%%%%%%%%%%%%%%%%%%%%%%%%%%%%%%%%%%%%%%%%

\section{CROSS-CORRELATION ANALYSIS}

%%%%%%%%%%%%%%%%%%%%%%%%%%%%%%%%%%%%%%%%%%%%%%%%%%%%%%%%%%%%%%%%%%%%%%%%%%%%%%%
\label{ccfana}

Quantitative estimates of lag time between the optical 
and near-infrared light curves are possible based on the
cross-correlation analysis commonly used in studies 
of multiple time series.  For example, the application of this type 
of analysis to spectroscopic monitoring observations of broad emission 
lines dates back more than a decade (for a review, 
see Peterson 2001).  

Let $C(t)$ and $I(t)$ be the irradiating ($V$ band) and 
responding ($K$ band) light curves at time $t$, 
respectively. Then they are related to each other via
\begin{equation}
I(t)=\int \Psi(\tau)\;C(t-\tau)d\tau \;\;,
\label{TF}
\end{equation}
where $\Psi(\tau)$ is the transfer function with respect 
to the lag time $\tau$.  This equation indicates that the 
transfer function represents the time-smeared response to 
a $\delta$-function outburst of incident radiation. 

As a function of lag time $\tau$, the cross-correlation 
function (CCF) and the auto-correlation function (ACF) are 
defined, respectively, as 
\begin{equation}
{\rm CCF}(\tau)=\int I(t)C(t-\tau)dt \;\;,
\label{ccf}
\end{equation}
and 
\begin{equation}
{\rm ACF}(\tau)=\int C(t)C(t-\tau)dt\;\;\;.
\label{acf}
\end{equation}
The ACF is therefore the cross-correlation of the incident 
light curve with itself.  Use of equation \ref{TF} to replace 
$I(t)$ in equation \ref{ccf} gives
\begin{equation}
{\rm CCF}(\tau)=\int \Psi(\tau'){\rm ACF}(\tau-\tau')d\tau'\;\;\;.
\label{ccf3}
\end{equation}
The CCF is therefore the convolution of the transfer function
with the ACF.  

The cross-correlation does not necessarily give an unambiguous 
measure of $\Psi(\tau)$, but simply quantifies the time delay 
$\tau$ between the two light curves, which approximately represents 
the location of the peak or the centroid of $\Psi(\tau)$.  
For intuitive interpretation of the CCF, we rewrite equation 
\ref{ccf} as follows: 
\begin{eqnarray}
{\rm CCF}(\tau) & = & \frac{1}{2} \int I(t)^2 dt \;+\; \frac{1}{2} 
\int C(t-\tau)^2 dt \nonumber \\
& - &\; \frac{1}{2}\int \left[ I(t)-C(t-\tau) \right]^2 dt\;\;\;.
\label{ccf4}
\end{eqnarray}
The first and second terms are nearly constant, if the interval 
of time integration is much larger than $\tau$.  Therefore, the 
third term, which is an integration of square difference between 
$I(t)$ and $C(t-\tau)$, determines the shape of CCF($\tau$). 
We see that the CCF becomes maximum when the third term becomes 
minimum.  Equation \ref{ccf4} supports our sight estimation of lag 
time by shifting either one of the irradiating and responding 
light curves with respect to time (Figures \ref{lc_ngc5548_VK} to 
\ref{lc_ngc7469_VK}), if two light curves are linearly related to
each other.

The CCF is equivalent to the correlation coefficient for the pairs 
of time series $(C(t-\tau),I(t))$. Thereby, practical calculation 
is given by
\begin{equation}
{\rm CCF}(\tau)=\frac{1}{N}\sum_{i=1}^{N}
\frac{[\,C(t_i-\tau)-\left<C\right>\,][\,I(t_i)-\left<I\right>\,]}
{\sigma_C\,\sigma_I} \;\;,
\label{ccf5}
\end{equation}
where $N$ is the number of pairs, and $\left<C\right>$ and 
$\left<I\right>$ are the averages of $C(t-\tau)$ and $I(t)$, 
respectively, and $\sigma_{C}$ and $\sigma_{I}$ are their standard 
deviations.  Note that this equation is valid only for the period 
during which $C(t-\tau)$ and $I(t)$ overlap in time.

\subsection{Interpolation Scheme}

The data points in equation \ref{ccf5} should be regularly spaced 
in time, because each point in one light curve must be paired with 
a corresponding point in the other.  For generating the 
pairs $(C(t-\tau),I(t))$, or equivalently $(V(t-\tau),K(t))$ in our 
case, linear interpolation of either one of the $V$ or $K$ light
curves has generally been used with small modifications 
if necessary (e.g., Gaskell \& Peterson 1987; White \& Peterson 1994). 

This interpolation method, however, sometimes does not 
provide a sufficient number of pairs for calculation, so that 
the shape of CCF is noisy and its peak difficult to detect.
Furthermore, this method does not guarantee a 
symmetrical ACF with respect to a lag time of zero, when the sampling 
of the data has a highly unbalanced weight for time. 

Here, we calculate CCFs based on different methods of interpolation, 
``bidirectionally interpolated'' (BI) CCFs and ``equally sampled'' (ES) CCFs. 
Let $\{V_1,V_2,...,V_N\}$ and $\{K_1,K_2,...,K_M\}$ be 
the fluxes of two observed light curves irregularly sampled at 
times $\{t_1,t_2,...,t_N\}$ and $\{t'_1,t'_2,...,t'_M\}$, respectively. 
Then, for the data pairs for which we calculate the standard linear correlation 
coefficient for lag time $\tau$, BI-CCFs combine two types of pairs, 
($V_{i},K_{\rm{int}}(t_{i}+\tau)$) and ($V_{\rm{int}}(t'_{j}-\tau),K_{j}$), 
where $V_{\rm{int}}$ and $K_{\rm{int}}$ are linearly interpolated fluxes 
from $\{V_{1},V_{2},...,V_{N}\}$ and $\{K_{1},K_{2},...,K_{M}\}$, 
respectively.  ES-CCFs collect the pairs of ($V_{\rm{int}}(t),K_{\rm{int}}
(t+\tau)$), where $t$ is sampled by equal interval $\Delta t$ independent 
of $\{t_{1},t_{2},...,t_{N}\}$ and $\{t'_{1},t'_{2},...,t'_{M}\}$.

These two methods provide symmetrical ACFs with respect to a 
lag time of zero. The BI-CCFs put weight on the observed data, 
yet are affected by the unbalanced sampling of observations 
concurrently. The ES-CCFs have contrary characteristics.
The problem in the interpolation, which occurs inevitably in both 
BI and ES methods, can be overcome by simulation of flux variations, 
as described below.

\begin{figure}
\epsscale{0.75}
\plotone{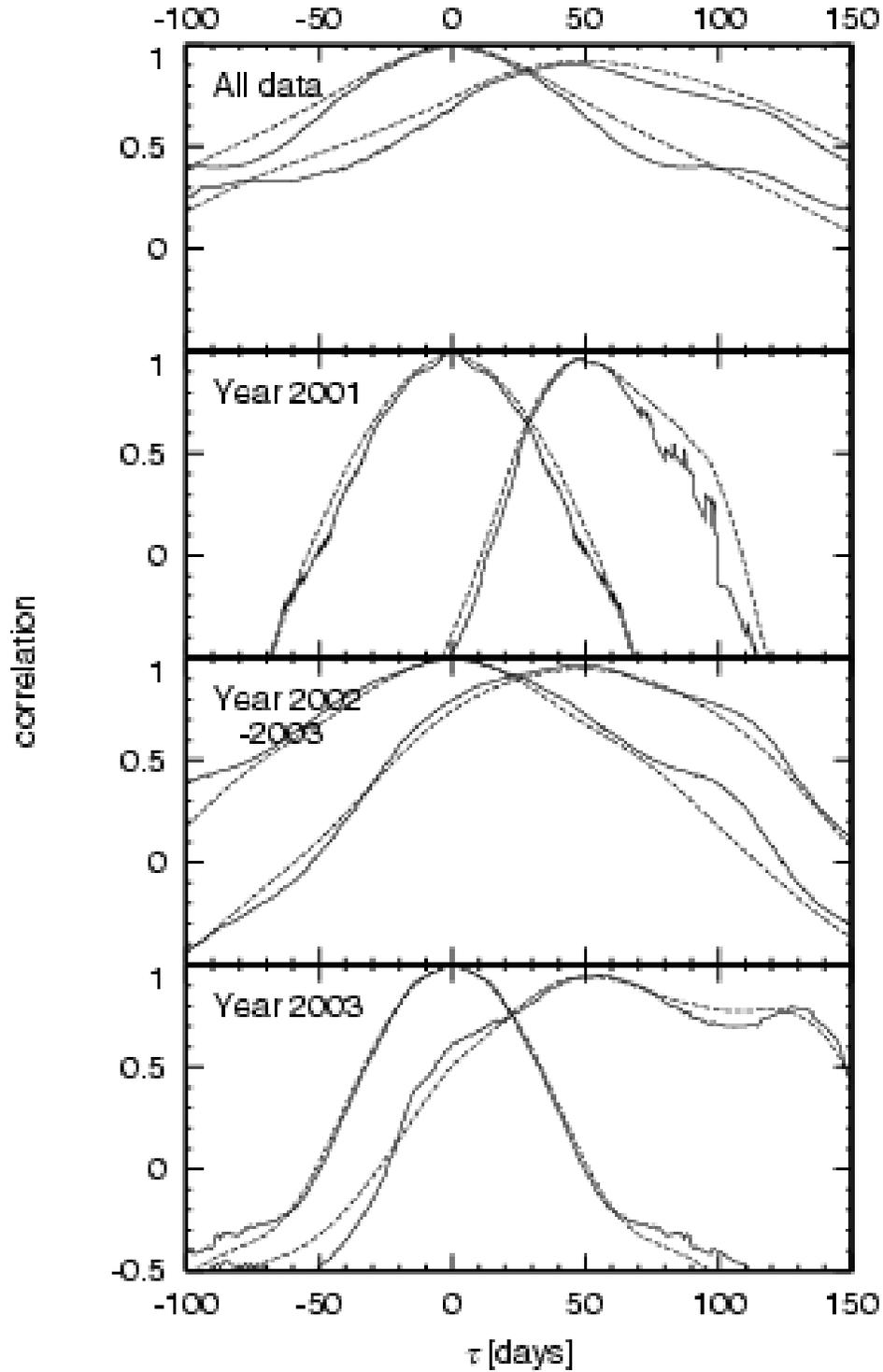}
\caption{
    CCFs and ACFs for various monitoring periods of
    NGC 5548.  The $V$ and $K$ light curves are used to calculate
    the CCFs.  The $V$ light curve is used to calculate the ACFs 
    that are symmetrical on either side of $\tau=0$.  The solid 
    and dashed lines are the results based on the BI and ES methods, 
    respectively.
\label{ccf_ngc5548}}
\end{figure}

\begin{figure}
\epsscale{0.75}
\plotone{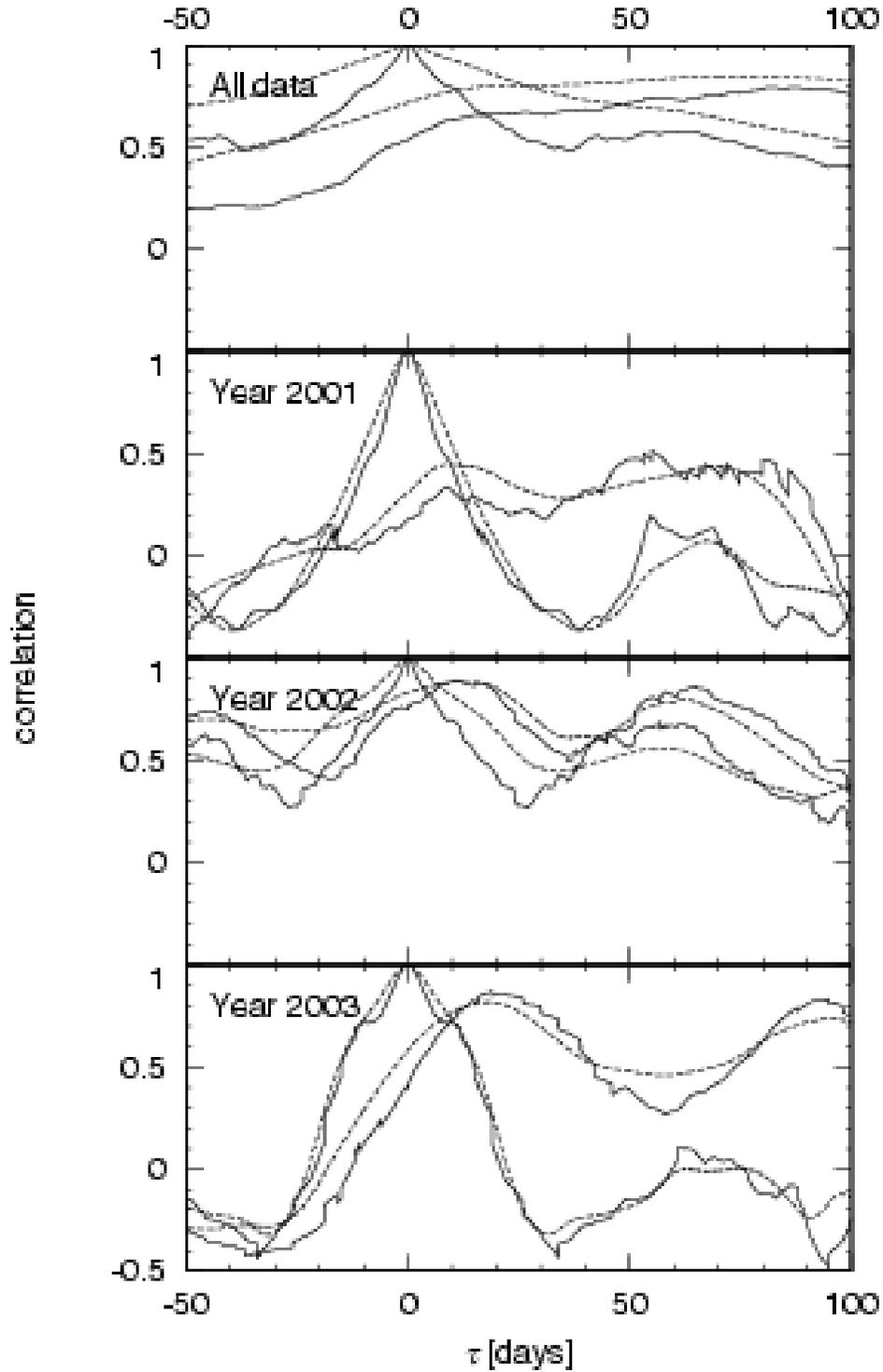}
\caption{
   CCFs and ACFs for various monitoring periods.  
   Same as Figure \ref{ccf_ngc5548}, but for NGC 4051.
    The CCFs are binodal with comparable peaks at 
    $\tau_{\rm peak}=10-20$ days and $50-100$ days. We
    regard the mode of shorter $\tau_{\rm peak}$ as 
    realistic one (see text).
\label{ccf_ngc4051}}
\end{figure}

\begin{figure}
\epsscale{0.75}
\plotone{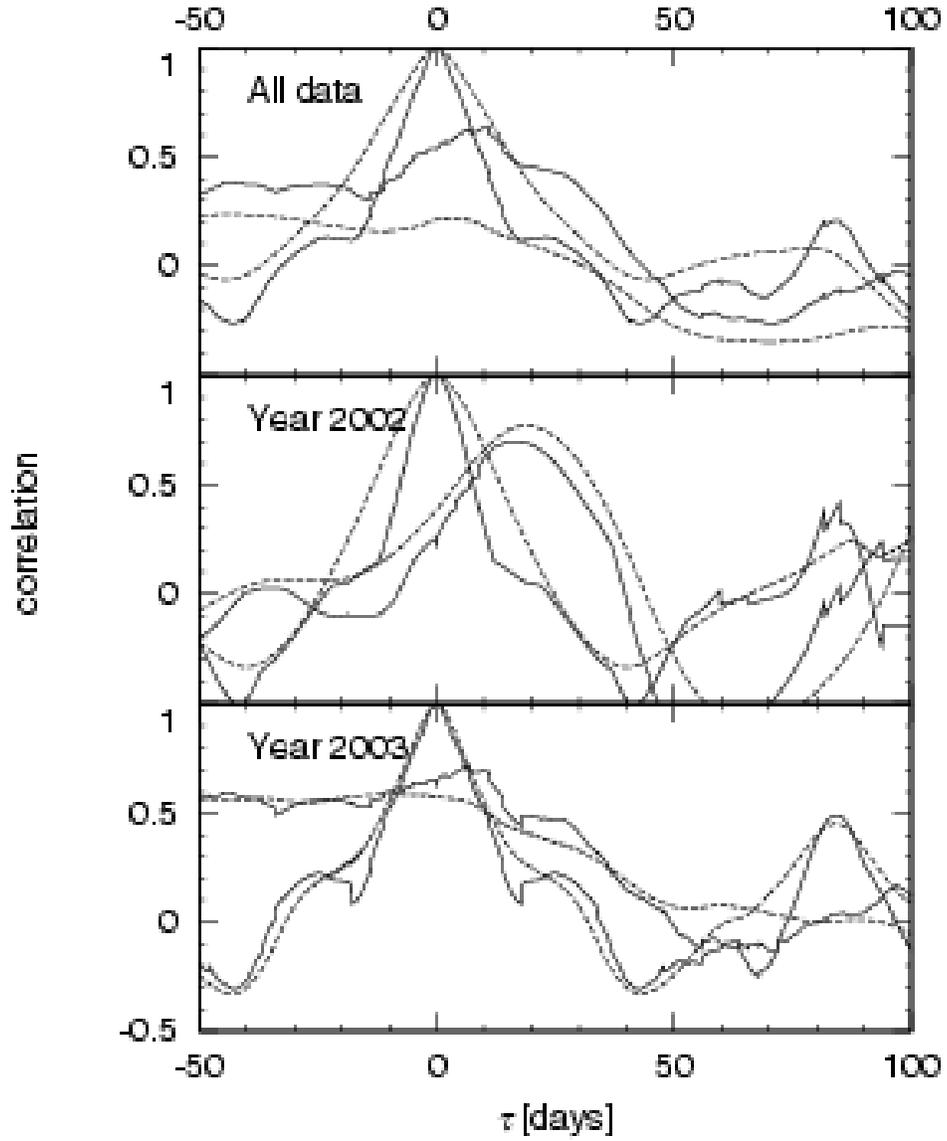}
\caption{
    CCFs and ACFs for various monitoring periods.  
    Same as Figure \ref{ccf_ngc5548}, but for NGC 3227.
    The CCFs on the top and bottom panels maintain a plateau 
    level with no clear peak below $\tau\approx 5-10$ days.}
\label{ccf_ngc3227}
\end{figure}

\begin{figure}
\epsscale{0.75}
\plotone{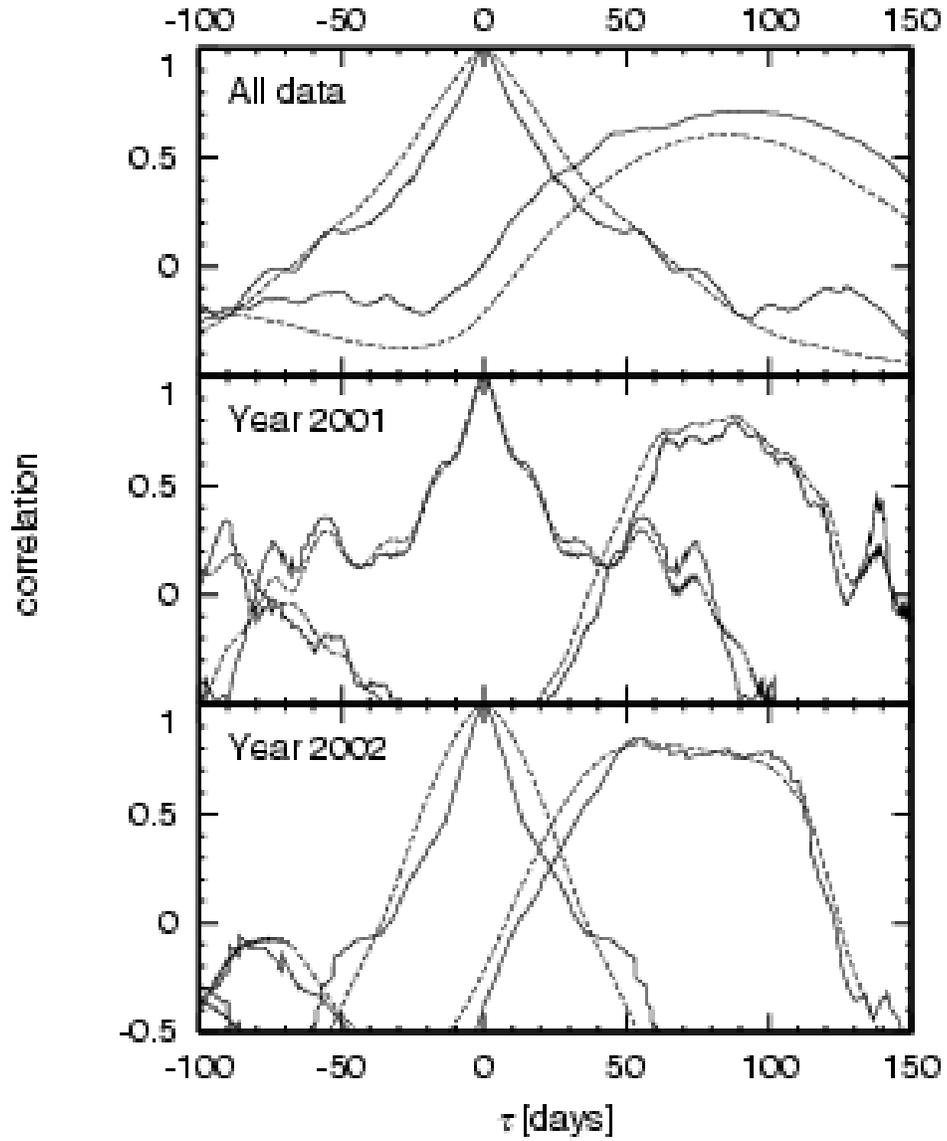}
\caption{
    CCFs and ACFs for various monitoring periods.  
    Same as Figure \ref{ccf_ngc5548}, but for NGC 7469.}
\label{ccf_ngc7469}
\end{figure}

Figures \ref{ccf_ngc5548} to \ref{ccf_ngc7469} show the CCFs and
ACFs for four target AGNs calculated by the two methods of BI and 
ES.
Table \ref{tp_VK} shows the cross-correlation results for several
subsets of the observed light curve data used in the calculation. 
The monitoring 
period for each subset, expressed in the modified Julian date (MJD), is
given in column 3.  The CCF peak position $\tau_{\rm peak}$ and the
CCF peak value $r_{\rm max}$ for the BI method are in columns 5 and 6, 
respectively, and those for the ES method are in columns 7 and 8.

\subsection{Uncertainty of Lag time}

Monte Carlo simulations are reliable for quantitative estimates 
of lag time.  Maoz \& Netzer (1989) introduced the basic scheme for 
this type of simulation.  A large number of artificial light curves 
are created for both irradiating and reprocessing sources, and 
their CCFs are calculated to determine the peak position
$\tau_{\rm{peak}}$ for each pair of light curves. 
The empirical distribution of $\tau_{\rm{peak}}$ is called 
`cross-correlation-peak-distribution (CCPD)'.  The uncertainty of 
lag time is determined from the CCPD.

In this technique, it is important to simulate the light curve 
such that the causes of the uncertainty should be adequately 
reflected in the uncertainty of lag time.  There are two major 
sources of uncertainty in lag time measurements.  One is flux 
errors in individual data points, and the other is the sampling 
gaps that often significantly underestimate the real variations.
In our photometric monitoring data, the former is not effective, 
because the photometric data have been obtained from homogeneous 
observation with very high S/N ratios, especially for nearby Seyfert 
galaxies.  On the other hand, the latter seriously affects the analysis, 
because there are inevitable gaps of three months due to solar conjunction 
and blanks of a few weeks due to bad weather conditions or facility 
maintenance.

Artificial light curves should be simulated in a less model-dependent 
way, based on the observed light curves, because we have a limited 
knowledge of the variability mechanism in the $V$ band and the 
transfer function from the $V$-band variations to the $K$-band 
variations.  Peterson et al. (1998) introduced a model-independent 
Monte Carlo method, called 
``flux randomization/random subset selection (FR/RSS)'', 
where the simulated light curve is based on the observed one.  

The FR method modifies the observed fluxes in each simulation 
by random Gaussian deviations based on the quoted error of each 
observed point, and accounts for the effects of flux-measurement 
uncertainties. 
The RSS method randomly extracts the same number of data points 
from the observed light curve, allowing for multiple extraction, to 
construct an artificial light curve. This method is then able to estimate 
the uncertainty due to the different time sampling of the data points.  
However, multiple extraction results in the removal of the observed 
data points at certain times in each simulation, which 
underestimates the real variation of the light curve. 

Here we introduce a new technique of realizing the light curves, 
which simulate the flux variations between the observed data points 
using stochastic processes characterized by the structure 
function (SF) from the observed light curves.  The SF is often 
used for the analysis of single time series (Simonetti et al. 1985). 
It measures the distribution of power over timescales for which the 
variations are correlated. A first-order structure function, for a 
series of flux measurements $f(t_i)$, $i=1,2,3,....$ is defined as
\begin{equation}
S(\tau)=\frac{1}{N(\tau)}\sum_{i<j}\left[ f(t_i)-f(t_j) \right]^2 \;\;,
\end{equation}
where the sum is over all pairs for which $t_j - t_i = \tau$, and 
$N(\tau)$ is the number of pairs.
The SF of most AGNs is given by
\begin{equation}
S(\tau) = S_{0}\;\tau^{\beta} \;\;,
\label{sf_pl}
\end{equation}
which increases up to some characteristic lag time of months or 
years.

At first, the mean square variation $\sigma^2(dt)$ in regard to 
interval $dt$ is determined based on the SF of the observed light 
curve.  In simulation of each light curve, an epoch during a 
sampling gap is randomly determined, and the flux is generated by 
random Gaussian deviations, which is equivalent to the FR method 
except for using the variability parameter 
$\sigma^2(dt)$ from the linearly interpolated flux from the 
observed data points. Once the point is achieved, it is regarded as 
an observed data point in the light curve, so that the realized 
data should have self-correlation seen in the observed light curves. 
Similar efforts are repeated until the mean sampling interval 
of the updated light curve decreases to one day.  The light curve, 
generated this way, is used in each CCF method, instead of using 
the interpolated fluxes in either of $V_{\rm{int}}$ or $K_{\rm{int}}$.

\subsection{Calculation and Simulation}
\label{ccf_calc}

\begin{figure}
\epsscale{0.9}
\plotone{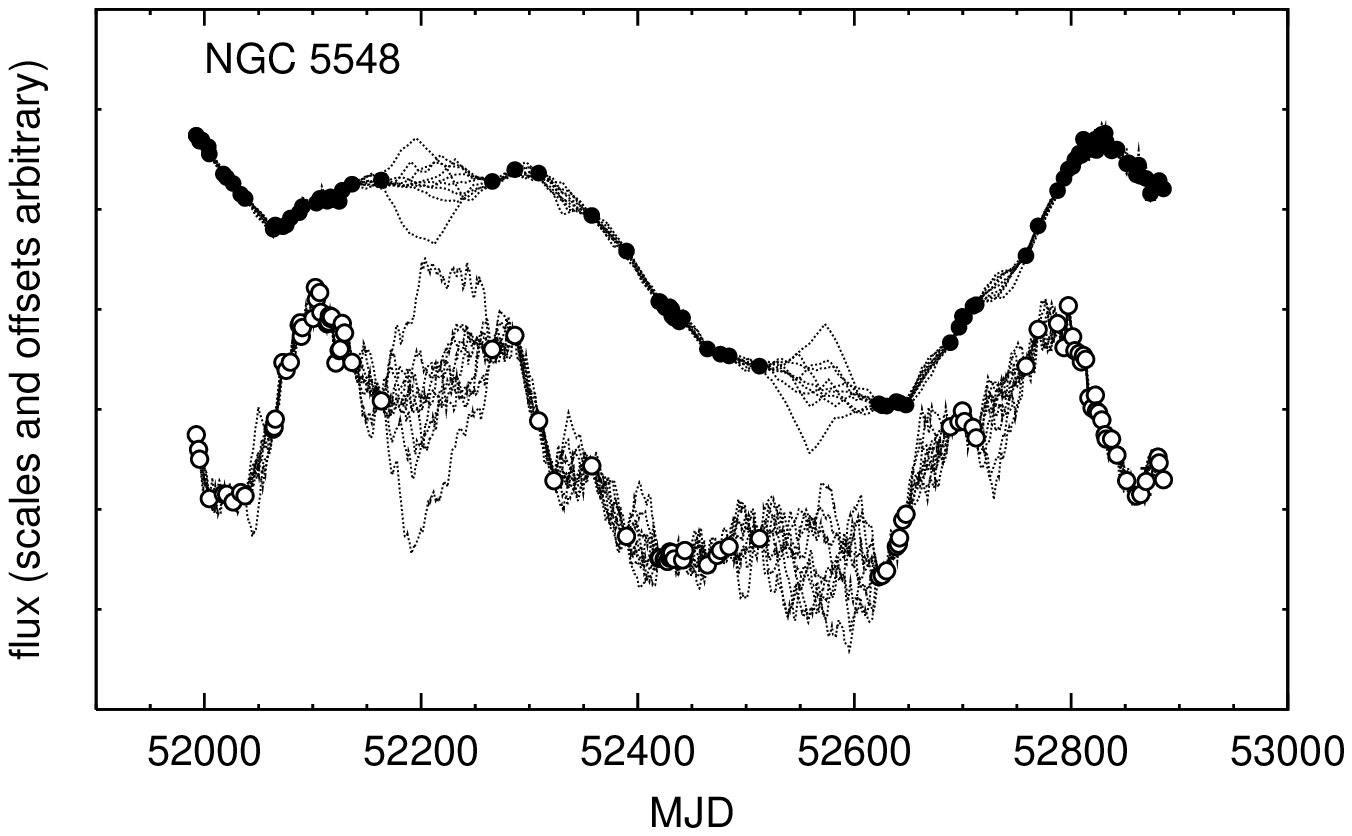}
\caption{Simulated $V$ and $K$ light curves of NGC 5548. Lower and 
      upper dotted lines correspond to $V$ and $K$, respectively. 
      Open and filled circles represent the observed $V$ and $K$ data 
      points, respectively.}
\label{simlc_ngc5548}

\clearpage

\epsscale{0.9}
\plotone{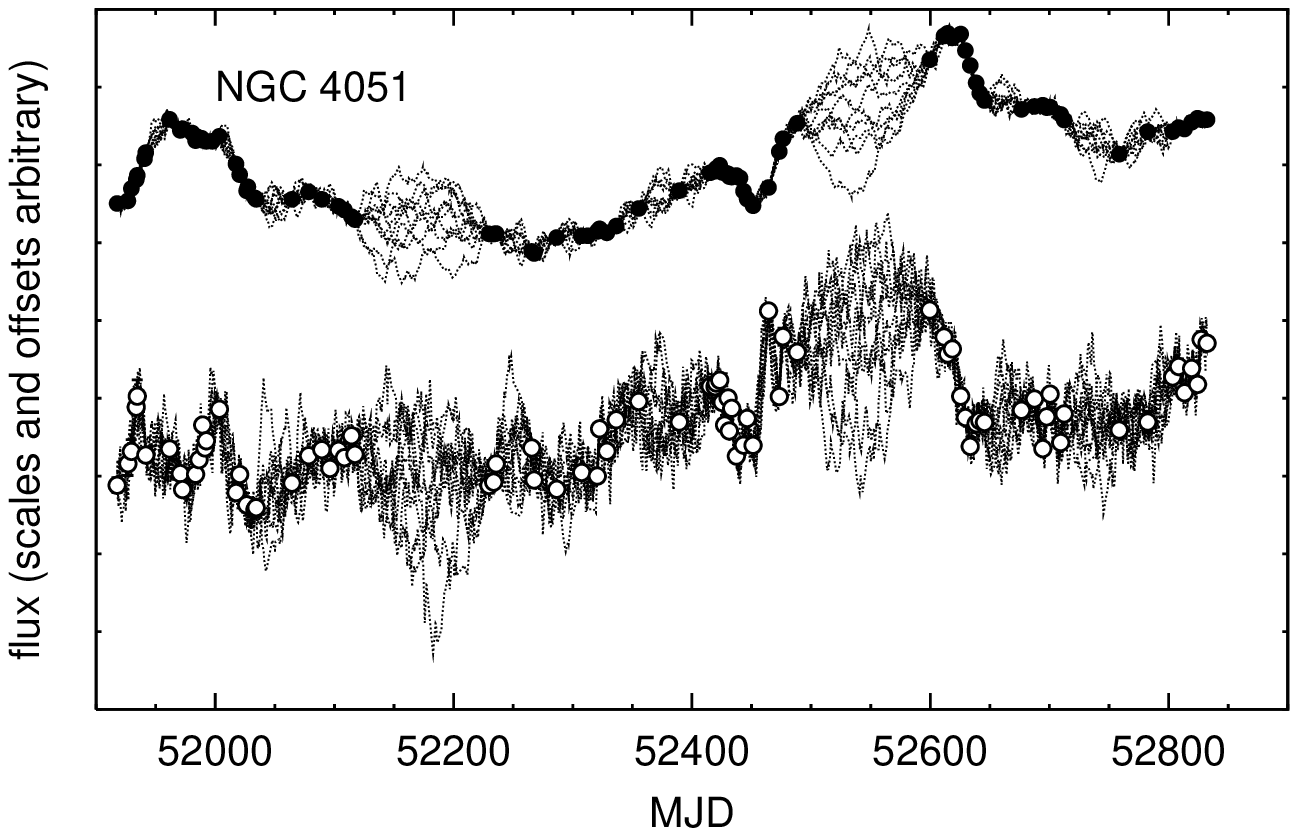}
\caption{Simulated $V$ and $K$ light curves of NGC 4051. Others
      are the same as Figure \ref{simlc_ngc5548}.}
\label{simlc_ngc4051}
\end{figure}

\begin{figure}
\epsscale{0.9}
\plotone{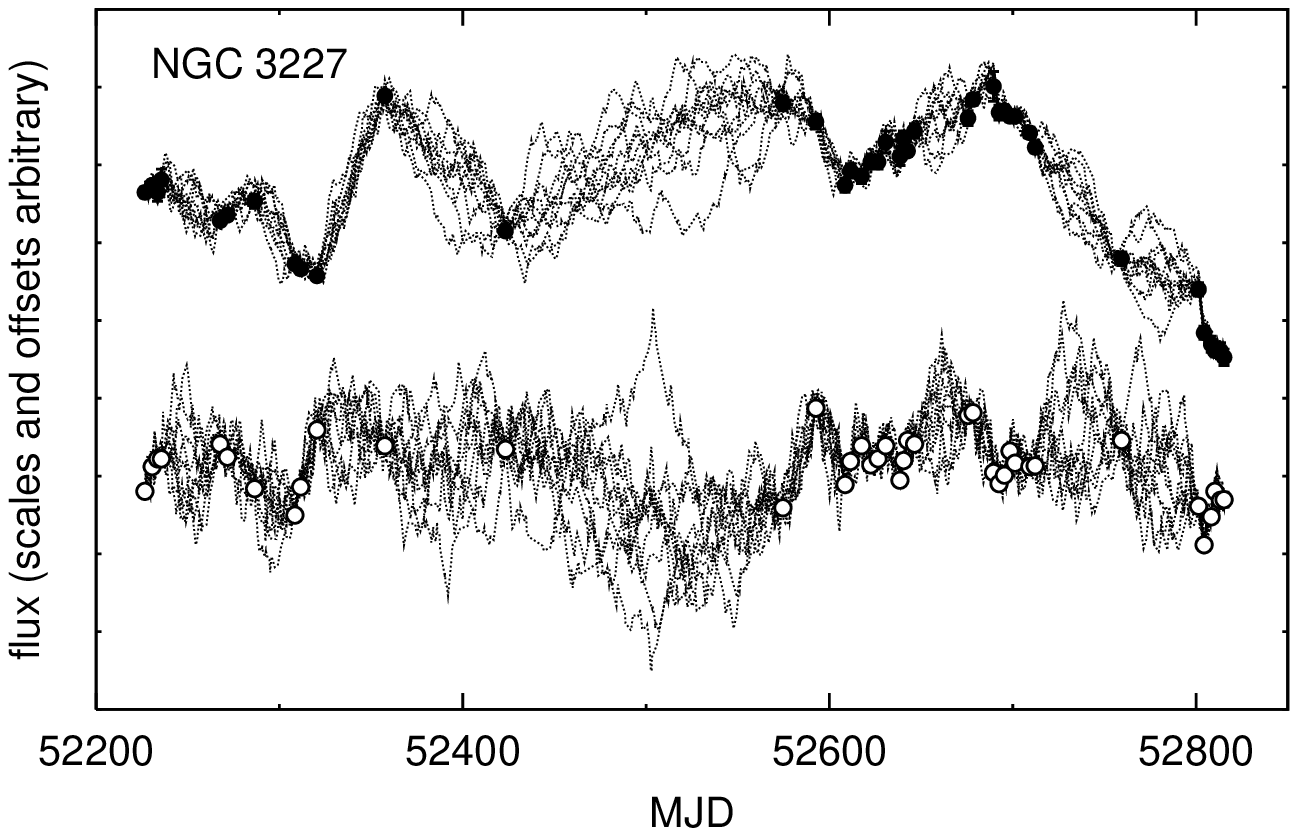}
\caption{Simulated $V$ and $K$ light curves of NGC 3227. Others
      are the same as Figure \ref{simlc_ngc5548}.}
\label{simlc_ngc3227}

\clearpage

\epsscale{0.9}
\plotone{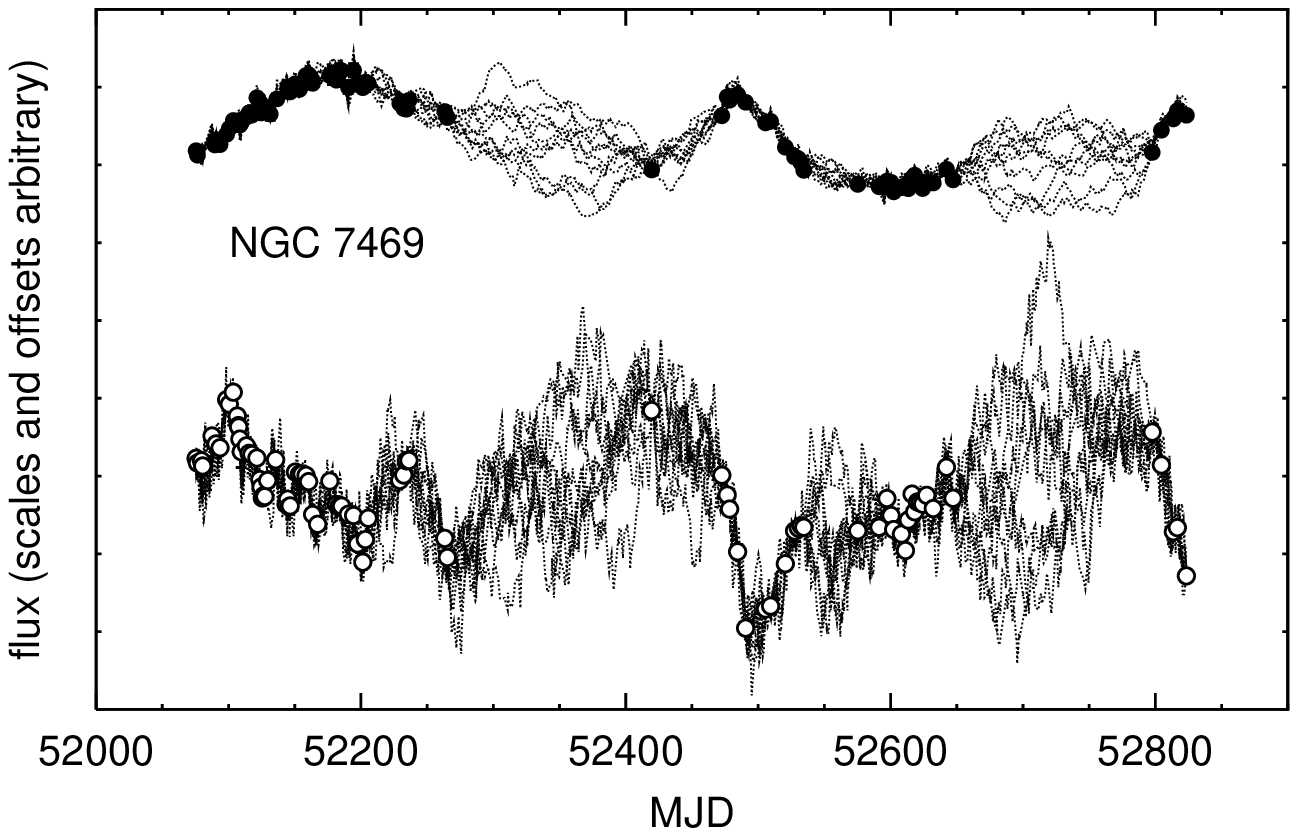}
\caption{Simulated $V$ and $K$ light curves of NGC 7469. Others
      are the same as Figure \ref{simlc_ngc5548}.}
\label{simlc_ngc7469}
\end{figure}

Table \ref{sf} shows the SFs fitted by equation \ref{sf_pl} using 
all the observed data in the $V$ and $K$ bands for each target 
AGN. The mean square variation $\sigma^{2}(dt)$ is determined 
directly from the observed SF.  However, in a particular case of 
the $K$ band for NGC 5548, it was determined as 
$\sigma^{2}(dt)=10\times(dt/60{\rm days})^{3}$ $({\rm mJy})^{2}$, 
by iteration until the simulated SF agrees with the observed SF.    
In this way we generated 1000 
pairs of artificial $V$ and $K$ light curves, some of which 
are shown in Figures \ref{simlc_ngc5548} to \ref{simlc_ngc7469}, 
together with the observed data.

The lag time was calculated for each subset of the monitoring period, 
because we are interested in whether phase-to-phase changes in lag 
time exist, and also because we know that the variation in the $K$ 
band does not always respond linearly to that in the $V$ band.  
The lag time for the entire monitoring period was also calculated, 
but this is only for reference. 

In principle, the lag time was calculated from the data taken
over a period between two adjacent solar conjunctions. Only for NGC 5548, 
we combined the second season of Year 2002 and a part of the third season 
of Year 2003, so that the lag time was calculated 
over a full period including the minimum luminosity state.

\begin{figure}
\epsscale{0.75}
\plotone{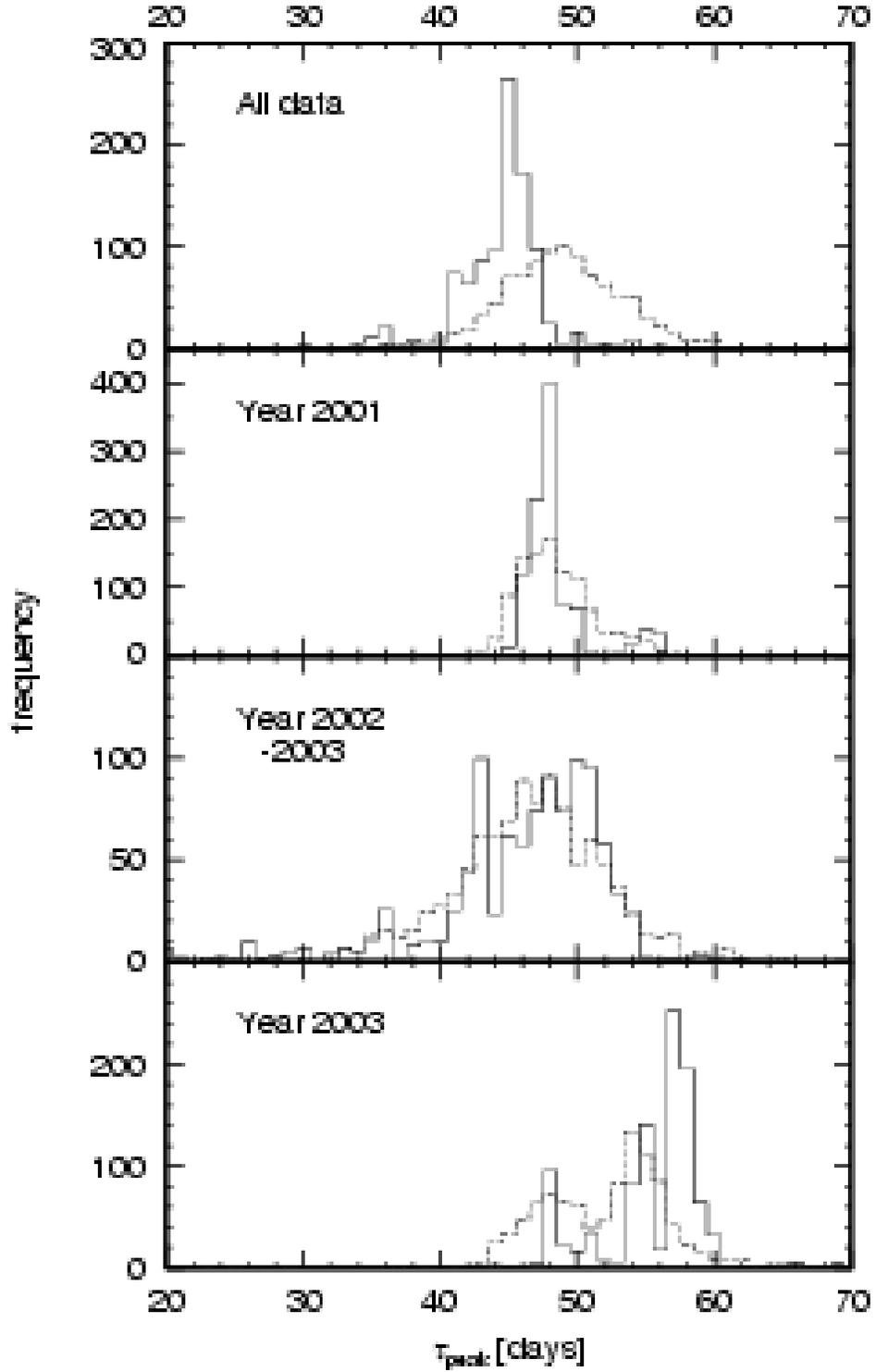}
\caption{
    CCPDs for various monitoring periods of NGC 5548. 
    The histogram shows the number of our simulation runs in each 
    bin of $\Delta\tau_{\rm peak}=2$ days.  The solid and dotted 
    lines are the results based on the BI and ES methods, respectively.}
\label{ccpd_ngc5548}
\end{figure}

\begin{figure}
\epsscale{0.75}
\plotone{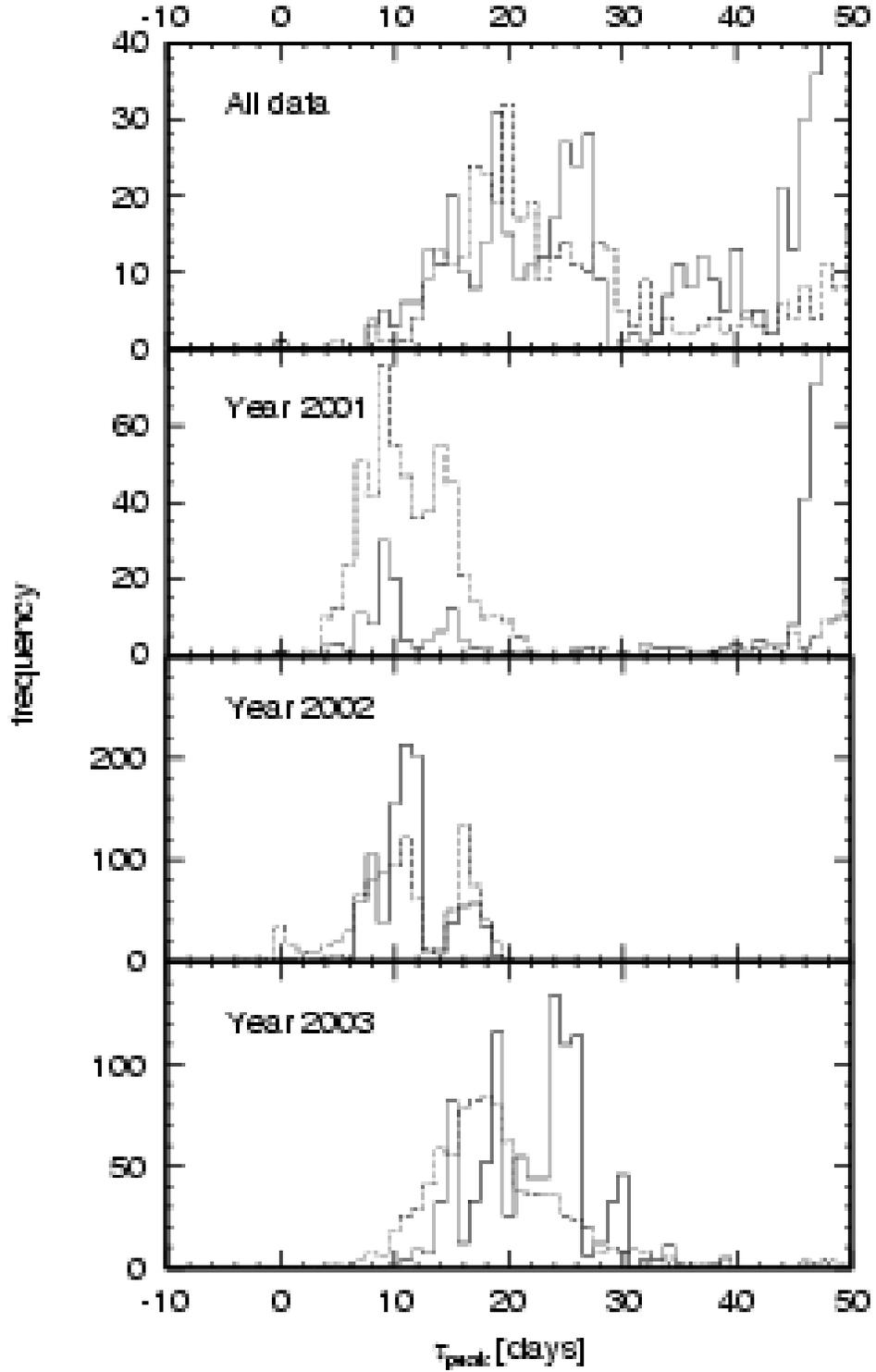}
\caption{
    CCPDs for various monitoring periods.
    Same as Figure \ref{ccpd_ngc5548}, but for NGC 4051.
    The lag time for the subset of all data as well as Year 2001 
    should be calculated from the histogram
    below $\tau_{\rm peak}\approx 42$ days (see text).}
\label{ccpd_ngc4051}
\end{figure}

\begin{figure}
\epsscale{0.75}
\plotone{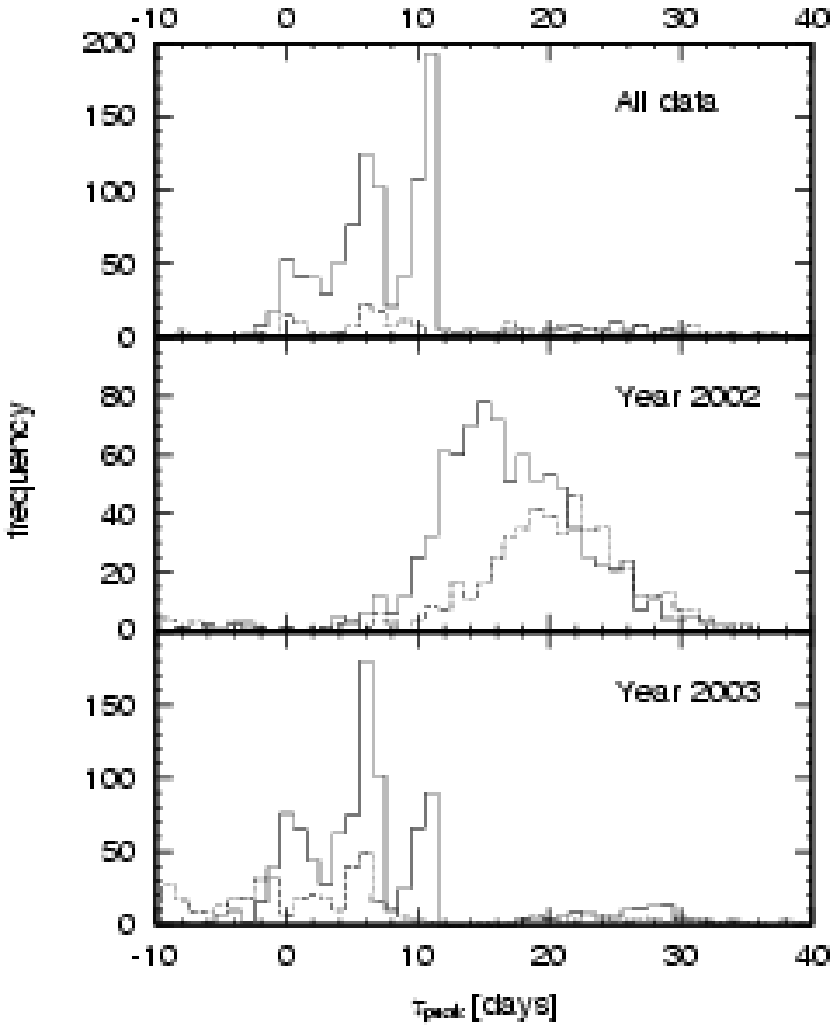}
\caption{
    CCPDs for various monitoring periods.
    Same as Figure \ref{ccpd_ngc5548}, but for NGC 3227.}
\label{ccpd_ngc3227}
\end{figure}

\begin{figure}
\epsscale{0.75}
\plotone{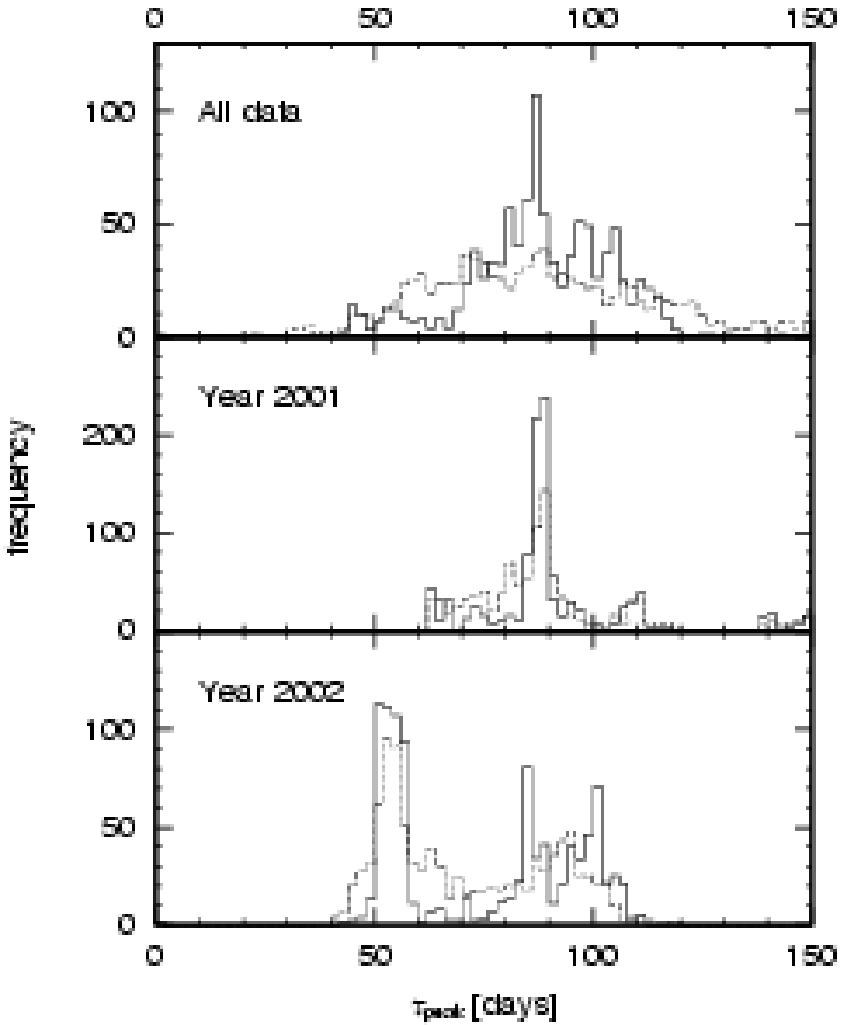}
\caption{
    CCPDs for various monitoring periods.
    Same as Figure \ref{ccpd_ngc5548}, but for NGC 7469.}
\label{ccpd_ngc7469}
\end{figure}

The CCPDs from Monte Carlo simulations are shown in Figures 
\ref{ccpd_ngc5548} to \ref{ccpd_ngc7469}, where the frequency 
distribution of $\tau_{\rm{peak}}$ over simulation runs is 
plotted as a histogram.  
Here, such runs of low CCF peaks with confidence levels of less than 
95 \% are excluded, where the confidence level is derived from the peak 
value of CCF and the number of observed data points in the overlapping 
period of the two time series shifted after by $\tau$.

Table \ref{lag_VK} shows the cross-correlation results from the 
simulated light curves.
The lag time was determined as a median of $\tau_{\rm{peak}}$,
or 50 \% fraction of the cumulative CCPD.  The uncertainty is 
measured by $\pm$34.1 \% fraction from the median, thus 
corresponding to 1$\sigma$ error for a normal statistical 
distribution.

The lag time for NGC 4051 needs careful interpretation.  
There are two comparable peaks in the CCF, located at 
$\tau_{\rm peak}=10-20$ days and $50-100$ days.  For the latter 
case, however, the major ascending and descending parts of the
$K$ light curve are shifted into the blank periods of the $V$ 
light curve, so that the high correlation may not be real.
The rapid $V$-band variations on short timescales of up to a month 
are not at all correlated with the $K$-band light curve if shifted 
backwards by  $50-100$ days.  According to equation \ref{ccf4}, 
the $V$ and $K$ variations of larger amplitude and longer timescale 
in the overlapping period tend to maximize the CCFs more effectively 
by chance, even if those of shorter timescale are in fact highly 
correlated.  It is very likely that the global flux variation for 
NGC 4051 contains information on some other effect of time shift in 
addition to the light-travel effect. Considering the above, the lag time 
shown in the table is calculated from the histogram below the dip at
$\tau_{\rm peak}\approx 42$ days. 

We detected the lag time for NGC 3227 for the first time,
though the uncertainty is not very small.  As seen 
from Figure \ref{lc_ngc3227_VK}, the optical variation of 
NGC 3227 is very rapid like that of NGC 4051.  The light curve data 
are undersampled because of some unfortunate gaps, especially 
in 2003.  More observations are needed to accurately measure 
the lag time for such an AGN with low-luminosity and rapid 
optical variation.

There is a possibility that the lag time may be underestimated, not representing 
the light-travel time from the central engine to the hot dust region.
The contribution of the $K$-band flux from the central engine, if any, 
would make the phase of near-infrared variations closer to the corresponding 
phase in the optical.  
In this case, the lag time should be systematically smaller than the 
light-travel time from the central engine to the hot dust 
region.  If the power-law optical component from the central engine 
is extended to the near-infrared wavelength region, its maximum 
contribution in the $K$-band flux is about 10 \%.  Subtraction of this 
component from the $K$-band flux observed on the same night would 
increase the lag time by $10-20$ \%.

%%%%%%%%%%%%%%%%%%%%%%%%%%%%%%%%%%%%%%%%%%%%%%%%%%%%%%%%%%%%%%%%%%%%%%%%%%%%%%%%

\section{DISCUSSION}

%%%%%%%%%%%%%%%%%%%%%%%%%%%%%%%%%%%%%%%%%%%%%%%%%%%%%%%%%%%%%%%%%%%%%%%%%%%%%%%%
\label{discus}

In most cases, the CCPDs are smooth and have clear peaks, 
and a lag between the $V$ and $K$ variations is detected 
(Figures \ref{ccpd_ngc5548} to \ref{ccpd_ngc7469}).  Generally, 
BI-CCFs tend to have a sharp jump, in contrast to ES-CCFs 
for which the pairs are sampled densely and constantly. Such a
jump appears close to the peak and affects the values of 
$\tau_{\rm peak}$.  We therefore use the ES-CCF result in 
Table \ref{lag_VK} as representing the lag time in following 
discussions. Note that the lag time for all data is only for
reference and is not shown in the plots.

\subsection{Relation between Infrared Lag and Central Luminosity}

The dust reverberation models, applied to previous observations of AGNs, 
have estimated an equilibrium temperature at the inner radius of the dust torus 
(Clavel, Wamsteker \& Glass 1989; Nelson 1996), which is consistent with the 
thermal emission in the near-infrared originating from the hot dust heated 
at its sublimation temperature in the range of $1300-2000$ K (Barvainis 1992).
As a result, there would be an inner hole of hot dust torus whose radius is 
proportional to a square root of central luminosity.  The lag time between 
UV/optical continuum variations and near-infrared ($\sim$2 $\mu$m) continuum 
variations should be the light-travel time from the center to the dust region.  

Our lag measurements of the four AGNs (NGC 5548, NGC 4051, NGC 3227, 
NGC 7469) are given in Table \ref{lag_VK}, and are plotted against
the absolute $V$ magnitude in Figure \ref{dtMv_all}, together with 
previous measurements. The filled symbols 
are our results, including our early result for NGC 4151 (Minezaki et al. 
2004). The others are taken from the literature (see Minezaki et al. 
2004 and references therein). It is evident from this figure that a good 
correlation of $\Delta t\propto L^{0.5}$, indicated earlier from $M_V=-23$ to
$-18$ by Minezaki et al. (2004), now extends down to $M_V=-15$, spanning
a range of 1000 times in luminosity.

In particular, we have expanded the range of $M_V$ faintwards by about three 
magnitudes by measuring, for the first time, the lag time of NGC 4051 which sits 
in the faint end of ``classical'' Seyfert nuclei. This indicates that AGNs of low 
luminosity might also have the dust torus expected from the dust reverberation models.

The good correlation between lag time and absolute $V$ magnitude strongly 
supports our idea of using $\Delta t\propto L^{0.5}$ as a distance indicator 
for AGNs (Kobayashi et al. 1998b; Yoshii 2002).  

Given a narrow range of flux variations throughout our monitoring 
period, we could not find clear differences among successive lag 
measurements for individual AGNs.  Observations covering larger 
flux variations are desirable for investigating how the hot dust 
would be redistributed against changes in the power of the central 
engine.

\subsection{Relation between Infrared Lag and Central Mass}

One of other fundamental quantities that characterize AGNs is the central 
mass, which could be regarded as the mass of a super-massive black hole. 
For the high-energy region, comparison of the characteristic frequency 
of X-ray variability power spectra of AGNs with that of black hole X-ray 
binaries has led to an interpretation that the size of the X-ray-emitting 
region would linearly scale with gravitational radius or gravitational 
mass, so that the X-ray variability has been proposed as a method of 
measuring the mass of AGNs (Hayashida et al. 1998; Uttley et al. 2002).  
For the UV or optical region, the optically thick and geometrically thin 
accretion disk is generally believed to be a source of UV and optical 
continuum from AGNs, and the standard model expects a radial temperature 
distribution in the disk that depends on both accretion rate and central 
mass (Frank, King, \& Raine 1992). Therefore, it is worthwhile to examine 
how the inner radius of the dust torus scales with central mass. 

Recent measurements of lag time and velocity dispersion for several
broad emission lines in NGC 5548 have given evidence for Kepler 
motion of the BLR gas (Peterson \& Wandel 1999).  Assuming that the 
virial theorem applies similarly to about three dozen Seyfert 
nuclei and quasars that have been targets of reverberation mapping 
observations for the BLR (e.g., Wandel, Peterson \& Malkan 1999; Kaspi 
et al. 2000; Peterson et al. 2004), their central masses have been 
estimated with a typical uncertainty of a factor of 2 or 3 over a range 
of $10^6-10^8 M_{\odot}$.

\begin{figure}
\epsscale{1.05}
\plotone{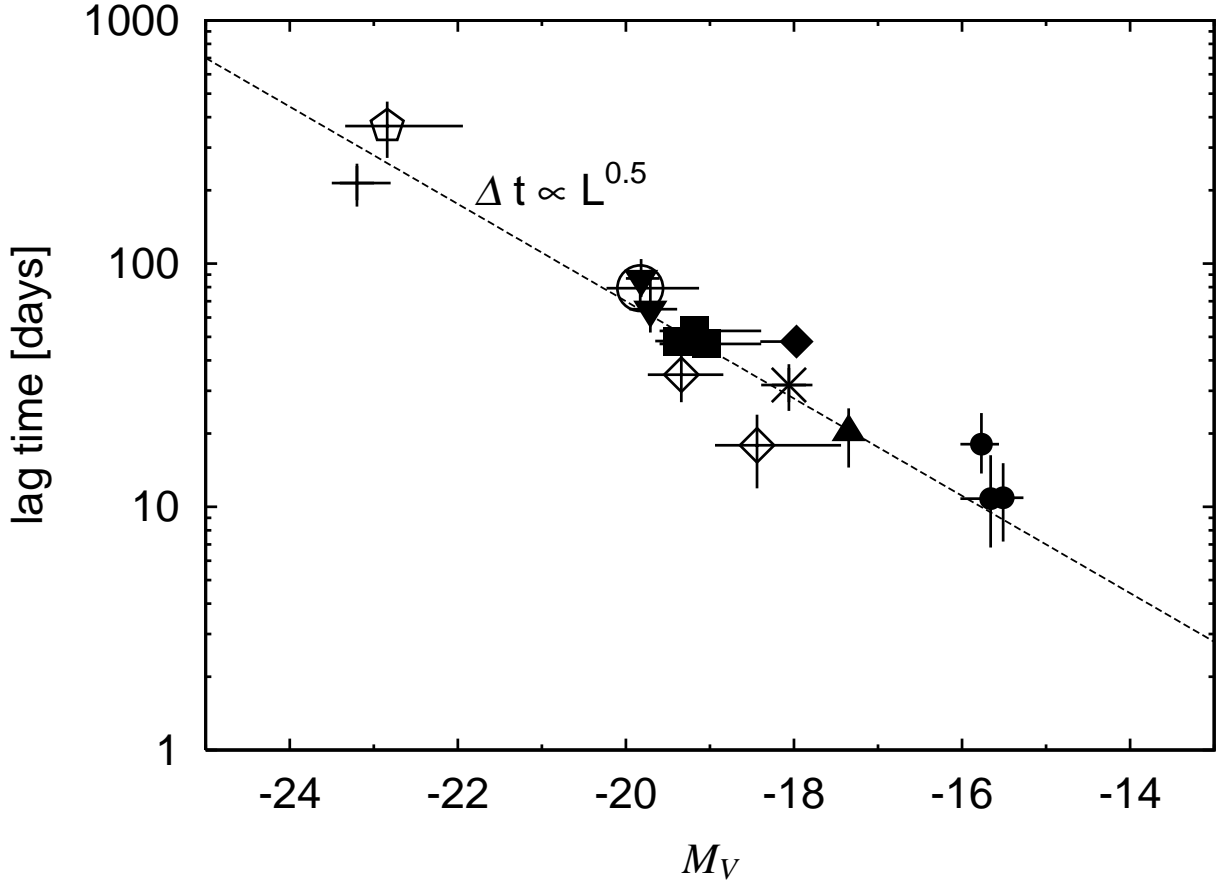}
\caption{Lag times between the UV/optical and $K$-band light curves 
    of AGNs, plotted against the absolute $V$ magnitude. 
    The filled symbols represent our results for NGC 5548 
    (squares), NGC 4051 (circles), NGC 3227 (triangle), and NGC 7469 
    (inverted triangles), including our earlier result for NGC 4151 
    (diamond, Minezaki et al. 2004). The others are data from the 
    literature for Fairall 9 (pentagon, Clavel, Wamsteker \& Glass 1989),
    GQ Comae (cross, Sitko et al. 1993), NGC 3783 (circle, Glass 1992),
    NGC 4151 (diamonds, Oknyanskij et al. 1999), and Mrk 744 
    (asterisk, Nelson 1996).
    Note that the horizontal bar in magnitude represents the 
    observed range of flux variation during the observation.}
\label{dtMv_all}
\end{figure}

\begin{figure}
\epsscale{1.05}
\plotone{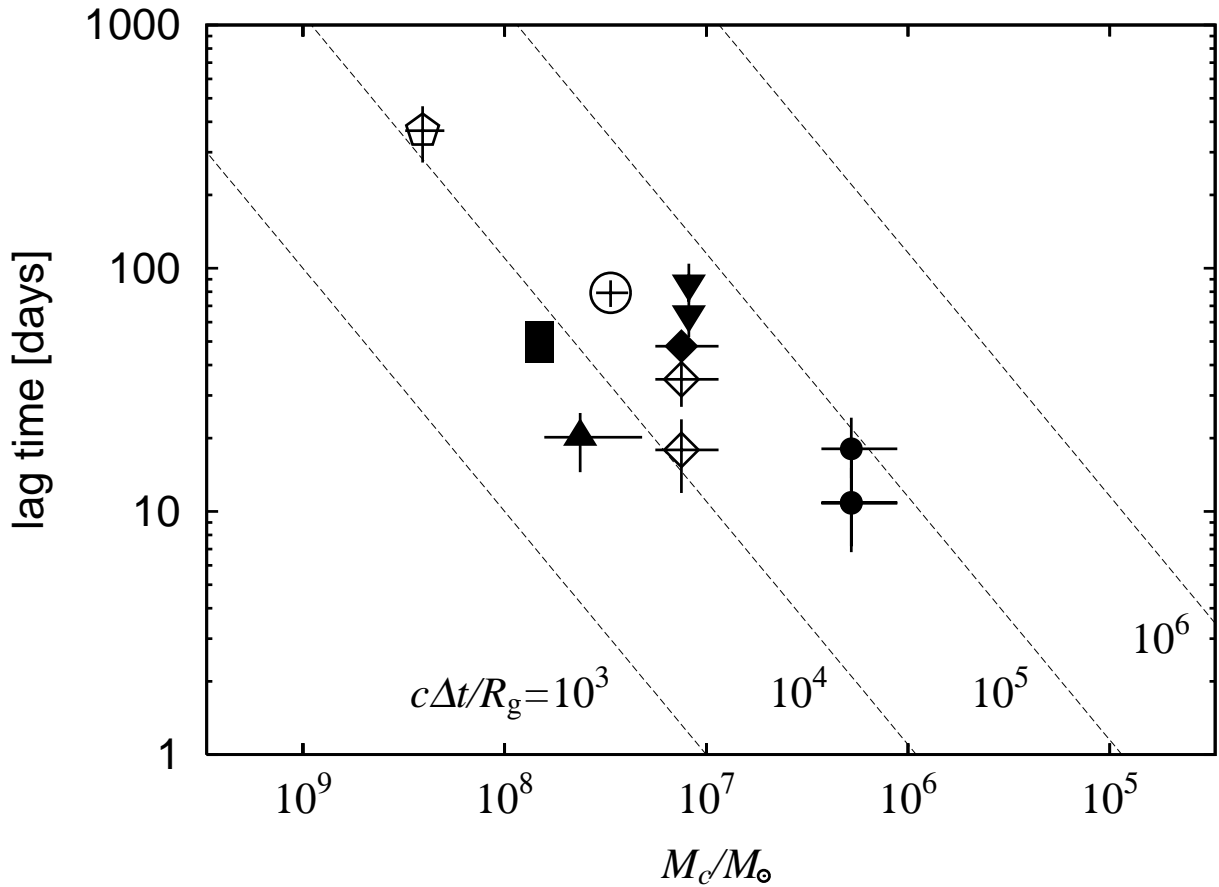}
\caption{Lag times for hot dust in Figure \ref{dtMv_all}, plotted
    against the central virial mass estimated from the reverberation 
    radii and widths of broad emission lines in individual AGNs
    (Peterson et al. 2004). 
    Each symbol indicates the same AGN as in Figure \ref{dtMv_all}. 
    The inclined lines for various values of $c\Delta t/R_g$ are shown,
    where $R_g=2GM_c/c^2$ is the gravitational radius and $M_c$ is the
    central virial mass.}
\label{dtMBH}
\end{figure}

Figure \ref{dtMBH} shows the infrared lag measurements plotted against 
the central masses that have been measured from reverberation radii and 
widths of broad emission lines in individual AGNs.  The meanings of 
filled and open symbols are the same as in Figure \ref{dtMv_all} for 
infrared lag measurements.  The central masses are taken from Peterson 
et al. (2004).  The inclined lines correspond to $c\Delta t/R_{g}=10^3$, 
$10^4$, $10^5$, and $10^6$, where $c\Delta t$ is the light-travel distance 
or inner radius of the dust torus, and $R_g=GM_c/c^2$ is the gravitational 
radius which scales linearly with central mass $M_c$.

Correlation between the central mass and the lag time (Figure \ref{dtMBH}) 
is weaker than between the central optical luminosity and the lag time
(Figure \ref{dtMv_all}).  Furthermore, the inner radius of dust torus 
$c\Delta t$ does not scale with gravitational radius $R_g$ that encloses 
the central mass. It is therefore evident that the inner radius of dust 
torus is not determined by the dynamics under the influence of a supermassive 
black hole at the center, but rather by the strength of UV/optical radiation 
from the central engine.  As a consequence, the scatter in Figure \ref{dtMBH} 
might be interpreted as the difference in mass-to-luminosity ratio that 
should be determined by the accretion rate and the radiative efficiency in 
accretion flow in individual AGNs.

\subsection{Relation of Dust Torus with Broad Line Region}

Confirmation of a dust torus outside the BLR not only verifies the 
unification scheme of AGNs, but also yields knowledge of physical
conditions in the central regions of AGNs for which we cannot yet obtain 
spatially resolved images.  Previously, Clavel, Wamsteker \& 
Glass (1989) measured both BLR and dust reverberation radii during 
the same monitoring campaign for a bright Seyfert 1 galaxy of 
Fairall 9, and concluded that the hot dust is outside the BLR.  We 
now examine whether this view is supported by our measurements 
of infrared lags.

\begin{figure}
\epsscale{0.95}
\plotone{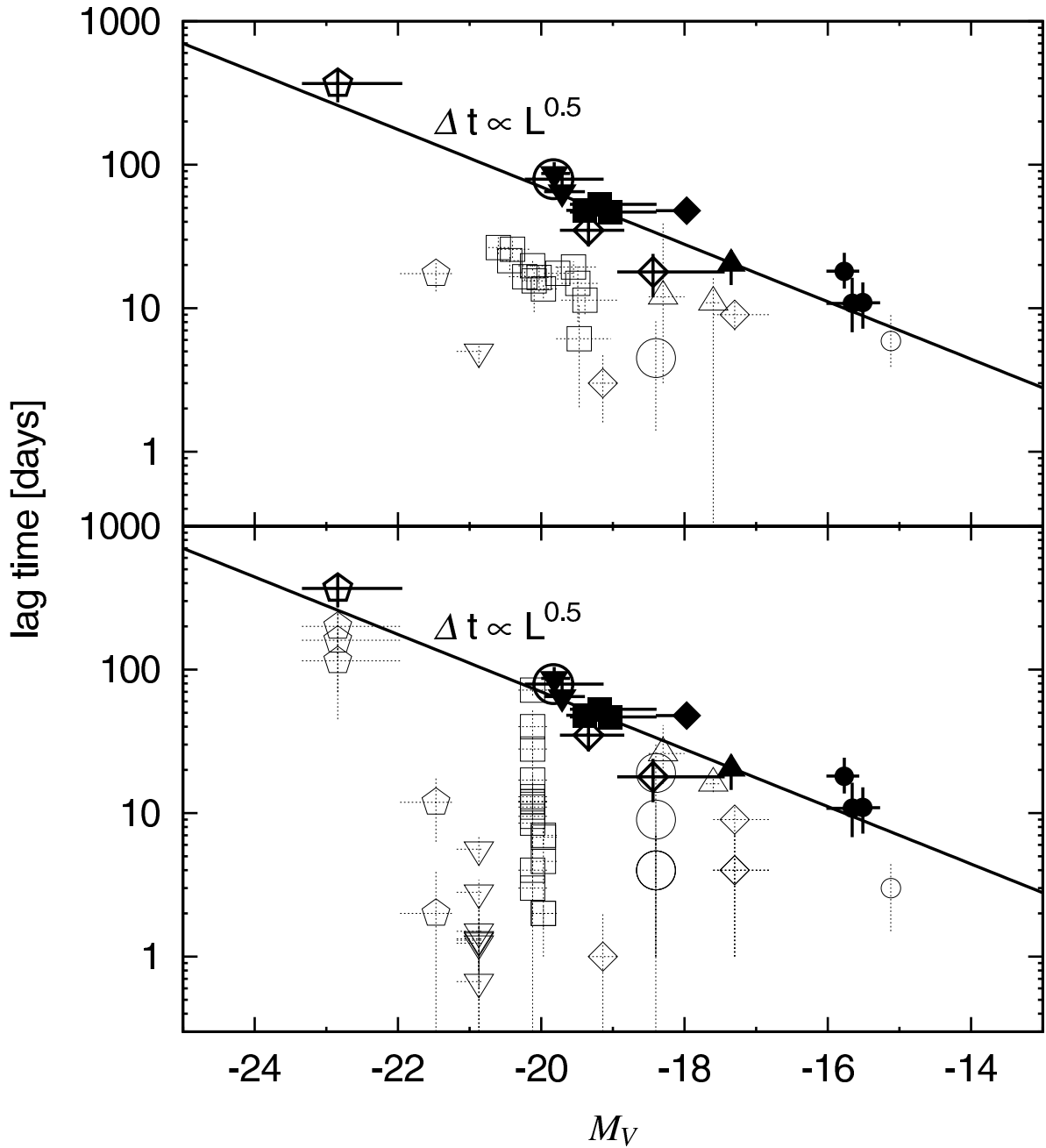}
\caption{Lag times plotted against the absolute $V$ magnitude
     of AGNs. 
     Filled and thick open symbols show the infrared 
     lags, where each symbol indicates the same AGN as in 
     Figure \ref{dtMv_all}. 
     Thin open symbols show the lag times of broad emission lines 
     for the corresponding AGNs in the literature. 
     {\it Top}: The lag times of only H$\beta$ line are plotted for broad
     emission lines. 
     {\it Bottom}: The lag times of other broad lines are plotted.}
\label{BLRdust}
\end{figure}

Figure \ref{BLRdust} shows our measurements of infrared lags 
(NGC 5548, NGC 4051, NGC 3227, NGC 7469; NGC 4151, Minezaki et al. 2004), 
together with previous measurements of broad-line lags in the literature 
(Fairall 9, Clavel, Wamstekar \& Glass 1989, Rodriguez-Pascual et al. 
1997, Peterson et al. 2004; NGC 3783, Reichert et al. 1994; NGC 7469, 
Wanders et al. 1997, Kriss et al. 2000, Collier et al. 1998; NGC 5548, 
Krolik et al. 1991, Peterson \& Wandel 1999, Dietrich et al. 1993, 
Korista et al. 1995; NGC 4151, Clavel et al. 1990, Maoz et al. 1991, 
Kaspi et al. 1996; NGC 3227, Winge et al. 1995, Onken et al. 2003; 
NGC 4051, Shemmer et al. 2003). The optical luminosities corresponding to 
the broad-line lags are averaged over the monitoring period in each item in the literature.

Each AGN has a range of lag time for the BLR depending on the broad emission 
line used (Korista et al. 1995; Peterson \& Wandel 1999; see also Figure \ref{BLRdust}).
In general, there is a tendency for broad emission lines of higher ionization 
to have smaller lag times, and for those of lower ionization to have larger lag times, 
which suggests that the BLR has a radially stratified ionization structure.  
Therefore, the BLR extends out to the light-travel distance corresponding to 
the largest lag time in each AGN.
We note that a correlation of $\Delta t \propto L^{0.7}$ has been suggested for
broad H$\beta$ emission lines for a wide range of central luminosity of AGNs including 
quasars brighter than $M_V \sim -23$ (Kaspi et al. 2000), although their lags for 
Seyfert galaxies show a fairly large scatter.

Figure \ref{BLRdust} shows that the lag times for broad lines are located 
below the boundary set by the infrared $\Delta t$ versus $M_V$ relation.
Some lags of broad emission lines are beyond the infrared lag in the 
same AGN. However, this occurs when such lags for the BLR were measured 
in a bright state, while the infrared lag for hot dust was 
measured in a faint state. Such lags for the
BLR, after being corrected for their luminosity dependence, become smaller than 
the infrared lag, as actually reported for NGC 5548 (Suganuma et al. 2004).

Thus, the outer radius of BLR nearly corresponds to the inner radius 
of the dust torus. In other words, it is suggested that the bulk of the emitting 
region of high-velocity gas is restricted inside the ``wall'' of the dust torus 
which is located at a distance determined by the continuum luminosity.  
Netzer \& Laor (1993) proposed that the dust in the narrow line-emitting gas
might separate the BLR from the NLR by suppressing the emission lines in the 
dust region, and that this might explain the large difference in observed properties 
of broad and narrow emission lines, such as line width and covering factor. 
The result presented in Figure \ref{BLRdust} supports their view that the
BLR size is determined by the dust sublimation radius.

\subsection{Optical Delay behind X-ray Variation in NGC 5548} 

It has been proposed that the UV/optical flux variation in AGNs 
originates in the thermal reprocessing of X-ray variability by the 
optically thick and geometrically thin accretion disk in its outer 
part, while the bulk of the steady component is generated by viscous processes in 
the accretion disk itself (e.g., Collin-Souffrin 1991). 
In this scheme, the variation of reprocessed component in the UV/optical 
flux should show a lag time of days or less behind the variation of X-rays 
released from the geometrically thick hot corona in the inner part of the accretion 
disk.  

For NGC 5548, Clavel et al. (1992) reported a UV lag of $\Delta t=0\pm 
6$ days behind the X-ray, by comparing the light curves in the hard 
X-ray ($2-10$ keV) and the UV (1350 \AA). It is not easy to detect a 
clear correlation between X-ray and UV/optical variations on short 
timescales, nor the lag time between them, because generally the amplitude 
of UV/optical variation on a short timescale is small and difficult to observe.

From observations of optical (5100 \AA) and hard X-ray ($2-10$ kev) 
fluxes of light curves spanning six years for NGC 5548, Uttley et al. (2003) 
reported a strong correlation in their long-timescale variation, while only weakly
restricting $\Delta t=0\pm15$ days. Since the amplitude of such optical variation is
found to exceed that of hard X-ray, they argued that X-ray reprocessing is not
a main driver of the long-timescale optical variation, at least for NGC 5548.

Thus, the problem yet to be answered is whether there is a clear correlation 
between X-ray and optical variations on short timescales.  As shown below, 
because of accurate photometry and frequent sampling of the $V$-band light 
curve data, we detected, for the first time, the day-basis optical variation having 
a small amplitude and a clear lag time behind the X-ray variation, which is well 
within a framework of reprocessing the X-ray flux into the optical flux. 

\begin{figure}
\epsscale{0.95}
\plotone{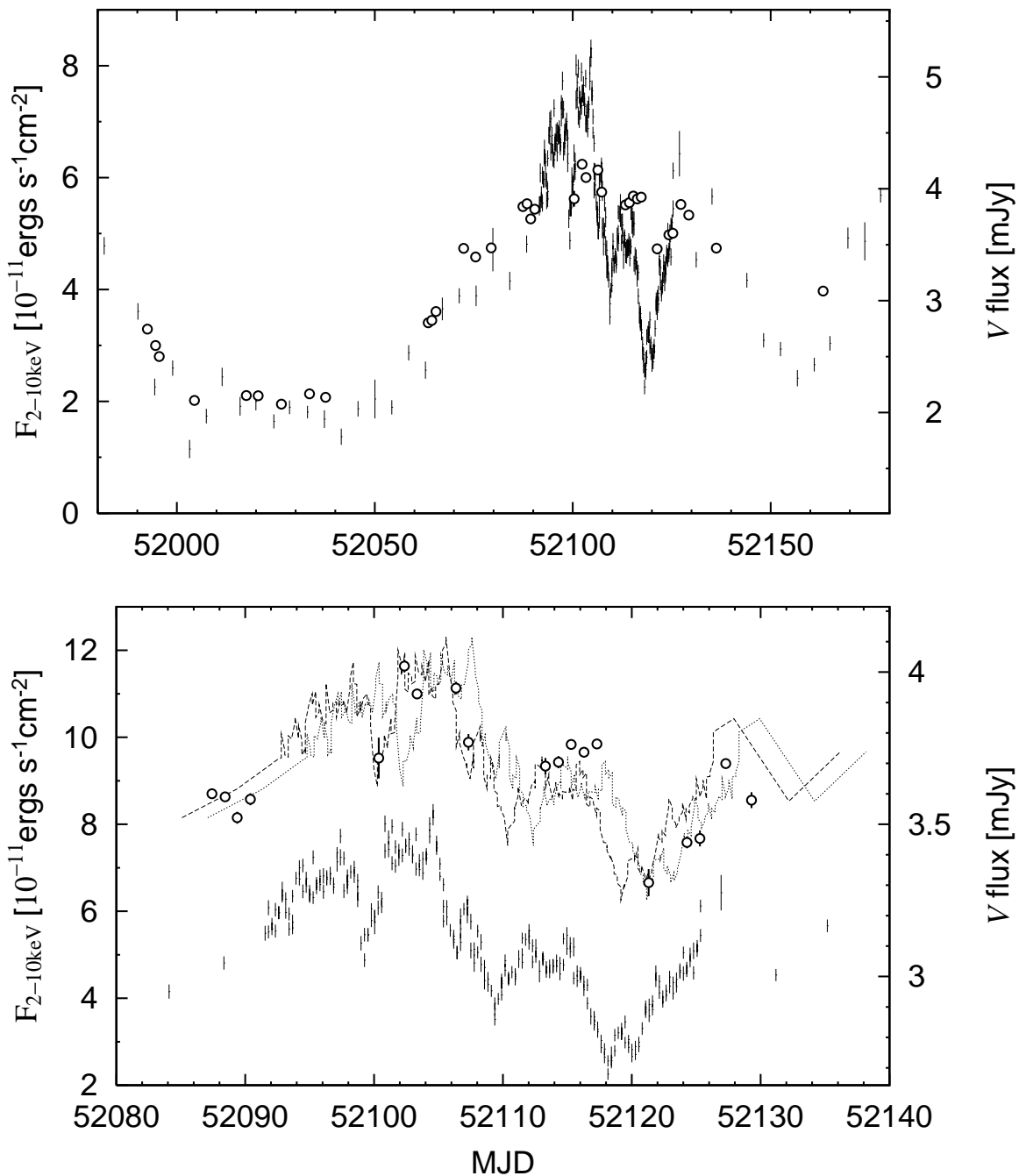}
\caption{{\it Top}: Observed light curves of NGC 5548 in the $V$ band 
    (open circles) and in the X-ray energy range of $2-10$ keV reported 
    by Uttley et al. (2003). 
    {\it Bottom}: Same as in the top panel, but only 
    for the active phase in 2001 (${\rm MJD} \approx 52085-52130$) when the 
    X-ray satellite (RXTE) observations were done intensively. 
    The dashed line is the RXTE light curve shifted arbitrarily in the
    vertical direction and by $+1$ days in the horizontal direction, 
    and the dotted line shifted by $+3$ days.  Note that the scale for 
    the optical flux is expanded in this panel.}
\label{ngc5548XV} 
\end{figure}

The top panel of Figure \ref{ngc5548XV} shows our $V$ light curve for 
NGC 5548 (Figure \ref{lc_ngc5548}) and the X-ray ($2-10$ keV) light 
curve obtained for the same period from the Rossi X-ray Timing Explorer 
(RXTE, Uttley et al. 2003).
These two light curves strongly correlate with each other, but the
amplitude of the optical variation is much smaller than that of the X-ray. 
The bottom panel shows a portion of the light curves for a period of 
${\rm MJD} \approx 52085-52130$ with an expanded vertical scale for the 
optical flux.  Although the optical data were not sampled as often 
as the RXTE, a day-basis correlated variation as well as an optical 
lag of one day or two behind the X-ray is clearly seen. 
Cross-correlation analysis gives a lag time of $\Delta t = 1.6_{-0.5}^{+1.0}$ 
days between the two light curves. A preliminary result was reported in
Suganuma et al. (2004).

The size of the X-ray emitting region in AGNs is considered to be as
small as a light-travel distance of hours, since a significant
fraction of X-ray flux varies on this timescale (Figure \ref{ngc5548XV}).
The observed correlated variability indicates that the optical emitting
region in NGC 5548 is located at a light-travel distance of one day or two
from the X-ray emitting region, and favors the thermal reprocessing of
X-ray variability by the optically thick accretion disk, rather than a
non-thermal process which emits both X-ray and UV/optical fluxes from
almost the same region.
On the other hand, the long-timescale X-ray and optical variations might be
generated by some other mechanisms, such as instabilities of the
accretion disk itself.

Our result also shows a geometrical relation between accretion disk and BLR.
As seen in Figure \ref{BLRdust}, the broad emission line of highest 
ionization in NGC 5548 gives the shortest lag of  a few days
 (Korista et al. 1995; Peterson \& Wandel 1999).  
The detected lag time of one or two days indicates that the UV/optical emitting 
region of the accretion disk is located in the innermost part of the BLR in NGC 5548.

%%%%%%%%%%%%%%%%%%%%%%%%%%%%%%%%%%%%%%%%%%%%%%%%%%%%%%%%%%%%%%%%%%%%%%%%%%%%%%%%

\section{SUMMARY}

%%%%%%%%%%%%%%%%%%%%%%%%%%%%%%%%%%%%%%%%%%%%%%%%%%%%%%%%%%%%%%%%%%%%%%%%%%%%%%%%

We carried out the most intense monitoring observations yet made in 
optical and near-infrared wavebands for four nearby Seyfert 1 galaxies, 
including NGC 5548, NGC 4051, NGC 3227, and NGC 7469.  Clear time-delayed 
response of the $K$-band flux variations to the $V$-band variations was 
detected for all the galaxies, which can be interpreted as light-travel 
time between the central engine and the inner edge of the dust torus.
We modified a cross-correlation analysis to measure the lag times with 
high reliability, and our infrared lag times are found to range from 10 
to 80 days, depending on the optical luminosity of the AGN.  Comparing 
our infrared lag measurements with other quantities (optical luminosity, 
central virial mass, and broad-line lag) based on our sample and others 
in the literature, we summarize our findings as follows:
\begin{enumerate}
\item The infrared lag time is tightly correlated with the optical 
  luminosity according to $\Delta t \propto L_{\rm opt}^{0.5}$ in
  agreement with the expected relation of dust reverberation. 
\item A large scatter between lag time and central virial mass 
  suggests that an inner radius of the dust torus is not directly 
  related to the dynamics of a supermassive black hole at the center.
\item Infrared lags place an upper boundary on similar lag 
  measurements of broad-emission lines in the literature.  This not 
  only supports the unified scheme of AGNs, but also implies a physical 
  transition from the BLR out to the dust torus that encircles the BLR. 
\item Our $V$ light curve for NGC 5548 shows a clear lag of one day or 
  two behind the X-ray light curve obtained from the RXTE satellite, 
  which indicates that the optical variation on short timescale originates in
  from the thermal reprocessing of X-rays by the optically thick accretion disk.
\end{enumerate}

Ongoing monitoring observations for AGNs in our sample, aided by the 
results from other samples and different wavelengths, would enable 
us to understand the size and structure of dust torus of AGNs, not 
only from their statistical properties but also from their time 
evolution.  Simultaneous reverberation studies of near-infrared 
continuum and broad emission lines of low ionization for several 
years would provide a clear view of the relation between dust torus 
and broad emission-line regions.

\acknowledgments

We thank Mamoru Doi and Kentaro Motohara for discussions and advice during 
the observations.  We also thank colleagues at the Haleakala Observatories 
for their help in facility maintenance. This research has been 
supported partly by the Grant-in-Aid of Scientific Research (10041110, 
10304014, 12640233, 14047206, 14253001, 14540223) and COE Research (07CE2002) 
of the Ministry of Education, Science, Culture and Sports of Japan, and has 
made use of the NASA/IPAC Extragalactic Database (NED) which is operated by 
the Jet Propulsion Laboratory, California Institute of Technology, under 
contract with the National Aeronautics and Space Administration.

\clearpage

\clearpage

\clearpage

\clearpage

\clearpage

\clearpage

\clearpage

\clearpage

\clearpage

\clearpage

\clearpage

\clearpage

\clearpage

\clearpage

\clearpage

\clearpage

\begin{deluxetable}{ccccrcc}
\tablecolumns{7}
\tablewidth{0pc}
\tablecaption{Characteristics of the target AGNs studied \label{objlst}}
\tablehead{
Object & $\alpha$(2000) & $\delta$(2000) & $m_V$\tablenotemark{a} & $cz$\tablenotemark{b} & 
$A_{V}$\tablenotemark{c} & $M_V$\tablenotemark{d}
}
\startdata
NGC5548 ...... & $14^{\rm h}$ $17^{\rm m}$ $59^{\rm s}.5$ & $+ 25^{\circ}$ $08^{'}$ $12^{''}.4$ & 15.3 & 5359 & 0.068 & $-19.2$ \nl
NGC4051 ...... & $12^{\;\;}$ $03^{\;\;\:}$ $09^{\;}.6$ & $+ 44^{\;\:}$ $31^{\;\:}$ $52^{\;\;}.8$ & 15.0 &  924 & 0.043 & $-15.7$ \nl
NGC3227 ...... & $10^{\;\;}$ $23^{\;\;\:}$ $30^{\;}.6$ & $+ 19^{\;\:}$ $51^{\;\:}$ $54^{\;\;}.0$ & 14.4 & 1480 & 0.075 & $-17.4$ \nl
NGC7469 ...... & $23^{\;\;}$ $03^{\;\;\:}$ $15^{\;}.6$ & $+ 08^{\;\:}$ $52^{\;\:}$ $26^{\;\;}.4$ & 14.5 & 4521 & 0.228 & $-19.8$ \nl
\enddata
\tablenotetext{a}{Average $V$-band nuclear magnitude from Table \ref{varstat}}
\tablenotetext{b}{Radial velocity in km s$^{-1}$ corrected for cosmic background from the NED database}
\tablenotetext{c}{Galactic $V$-band extinction from the NED database}
\tablenotetext{d}{Absolute $V$-band nuclear magnitude assuming 
$H_0=70$ km s$^{-1}$Mpc$^{-1}$}
\end{deluxetable}

\begin{deluxetable}{clccc r@{}c@{}l r@{}c@{}l}
\tablecolumns{11}
\rotate
\tablewidth{0pc}
\tablecaption{Apparent magnitudes of reference stars in the optical bands 
\label{refmagopt}}
\tablehead{
 Object & Star & $\alpha$(2000) & $\delta$(2000) & $U$ (mag) & \multicolumn{3}{c}{$B$ (mag)} & \multicolumn{3}{c}{$V$ (mag)}
}
\startdata
NGC 5548 & refA  & $14^{\rm h}$ $17^{\rm m}$ $58^{\rm s}.9$ & $+25^{\circ}$ $05^{'}$ $34^{''}$ & -                  & 14.422&$\pm$&0.006 (19) & 13.782&$\pm$&0.003 (37) \\
         & refB1       & $14^{\;\;}$ $17^{\;\;\:}$ $14^{\;\:}.7$ & $+25^{\;\:}$ $19^{\;\:}$ $22^{\;\;}$ & -                  & 13.742&$\pm$&0.004 (19) & 13.129&$\pm$&0.002 (37) \\
NGC 4051 & refA  & $12^{\;\;}$ $04^{\;\;\:}$ $13^{\;\:}.4$ & $+44^{\;\:}$ $55^{\;\:}$ $03^{\;\;}$ & -                  & 13.568&$\pm$&0.007 (13) & 12.636&$\pm$&0.003 (27) \\
         & refB2       & $12^{\;\;}$ $03^{\;\;\:}$ $30^{\;\:}.2$ & $+44^{\;\:}$ $16^{\;\:}$ $46^{\;\;}$ & -                  & 14.472&$\pm$&0.005 (13) & 13.874&$\pm$&0.006 (27) \\
NGC 3227 & refA  & $10^{\;\;}$ $23^{\;\;\:}$ $13^{\;\:}.6$ & $+19^{\;\:}$ $56^{\;\:}$ $25^{\;\;}$ & -                  & 13.394&$\pm$&0.003 ( 7) & 12.771&$\pm$&0.003 (12) \\
         & refB         & $10^{\;\;}$ $24^{\;\;\:}$ $02^{\;\:}.9$ & $+19^{\;\:}$ $46^{\;\:}$ $31^{\;\;}$ & -                  & 12.506&$\pm$&0.003 ( 7) & 11.959&$\pm$&0.003 (12) \\
NGC 7469 & refA  & $23^{\;\;}$ $03^{\;\;\:}$ $07^{\;\:}.1$ & $+08^{\;\:}$ $34^{\;\:}$ $30^{\;\;}$ & 14.21$\pm$0.01 (6) & 13.918&$\pm$&0.003 ( 4) & 13.180&$\pm$&0.003 (18) \\
         & refB         & $23^{\;\;}$ $02^{\;\;\:}$ $23^{\;\:}.4$ & $+09^{\;\:}$ $06^{\;\:}$ $41^{\;\;}$ & 14.43$\pm$0.01 (6) & 13.891&$\pm$&0.003 ( 4) & 12.998&$\pm$&0.003 (18) \\
\enddata
\tablecomments{The number of photometric calibrations for each star and band is given in the parentheses.}
\end{deluxetable}

\begin{deluxetable}{clcc l@{}c@{}l@{}l l@{}c@{}l@{}l l@{}c@{}l}
\tablecolumns{15}
\rotate
\tablewidth{0pc}
\tablecaption{Apparent magnitudes of reference stars in the near-infrared bands  \label{refmagir}}
\tablehead{
 Object & Star & $\alpha$(2000) & $\delta$(2000) & \multicolumn{4}{c}{$J$ (mag)} & \multicolumn{4}{c}{$H$ (mag)} & \multicolumn{3}{c}{$K$ (mag)}
}
\startdata
NGC 5548 & refA  & $14^{\rm h}$ $17^{\rm m}$ $58^{\rm s}.9$ & $+25^{\circ}$ $05^{'}$ $34^{''}$  & &-& &                              & 12.227&$\pm$&0.004&\hspace{1mm}(3)  & 12.196&$\pm$&0.006 ( 9)\\
         & refB1       & $14^{\;\;}$ $17^{\;\;\:}$ $14^{\;\:}.7$ & $+25^{\;\:}$ $19^{\;\:}$ $22^{\;\;}$  & &-& &                              & &-& &                                      & 11.531&$\pm$&0.004 ( 9)\\
         & refB2       & $14^{\;\;}$ $17^{\;\;\:}$ $12^{\;\:}.2$ & $+25^{\;\:}$ $19^{\;\:}$ $33^{\;\;}$  & &-& &                              & 13.072&$\pm$&0.009&\hspace{1mm}(3)  & &-&\\
NGC 4051 & refA  & $12^{\;\;}$ $04^{\;\;\:}$ $13^{\;\:}.4$ & $+44^{\;\:}$ $55^{\;\:}$ $03^{\;\;}$  & 10.606&$\pm$&0.004&\hspace{1mm}(4) & 10.06&$\pm$&0.01  &\hspace{1mm}(1)    & 9.948&$\pm$&0.004 (13)\\
         & refB1       & $12^{\;\;}$ $03^{\;\;\:}$ $34^{\;\:}.0$ & $+44^{\;\:}$ $16^{\;\:}$ $50^{\;\;}$  & 10.545&$\pm$&0.004&\hspace{1mm}(4) & &-& &                    & 10.243&$\pm$&0.003 (13)\\
         & refB2       & $12^{\;\;}$ $03^{\;\;\:}$ $30^{\;\:}.2$ & $+44^{\;\:}$ $16^{\;\:}$ $46^{\;\;}$  & &-& &                              & 12.34&$\pm$&0.01  &\hspace{1mm}(1)    & &-&\\
NGC 3227 & refA  & $10^{\;\;}$ $23^{\;\;\:}$ $13^{\;\:}.6$ & $+19^{\;\:}$ $56^{\;\:}$ $25^{\;\;}$  & 11.483&$\pm$&0.005&\hspace{1mm}(5) & 11.143&$\pm$&0.004&\hspace{1mm}(9)  & 11.092&$\pm$&0.004 (13)\\
         & refB         & $10^{\;\;}$ $24^{\;\;\:}$ $02^{\;\:}.9$ & $+19^{\;\:}$ $46^{\;\:}$ $31^{\;\;}$  & 10.806&$\pm$&0.005&\hspace{1mm}(5) & 10.508&$\pm$&0.004&\hspace{1mm}(9)  & 10.459&$\pm$&0.004 (13)\\
NGC 7469 & refA  & $23^{\;\;}$ $03^{\;\;\:}$ $07^{\;\:}.1$ & $+08^{\;\:}$ $34^{\;\:}$ $30^{\;\;}$  & 11.74&$\pm$&0.01 &\hspace{1mm}(2)  & 11.366&$\pm$&0.010&\hspace{1mm}(4)  & 11.308&$\pm$&0.006 ( 9)\\
         & refB         & $23^{\;\;}$ $02^{\;\;\:}$ $23^{\;\:}.4$ & $+09^{\;\:}$ $06^{\;\:}$ $41^{\;\;}$  & 11.22&$\pm$&0.01 &\hspace{1mm}(2)  & 10.778&$\pm$&0.010&\hspace{1mm}(4)  & 10.696&$\pm$&0.004 ( 9)\\
\enddata
\tablecomments{The number of photometric calibrations for each star and band is given in the parentheses. 
The uncertainties of the photometric standard stars, typically about 0.01 mag, are not included in the calibrations. }
\end{deluxetable}

\begin{deluxetable}{c c c c c c c }
\tablecolumns{19}
\tablewidth{0pc}
\tablecaption{Host galaxy flux\tablenotemark{a} in the $\phi=8''.3$ aperture \label{galfx}}
\tablehead{
 Object & $U$ & $B$ & $V$ & $J$ & $H$ & $K$
}
\startdata
NGC 5548 ...... &   - & 1.4 & 3.7 &  - & 20 & 16\\
NGC 4051 ...... &   - & 4.7 & 8.6 & 41 & 47 & 39\\
NGC 3227 ...... &   - & 3.0 & 6.4 & 50 & 66 & 60\\
NGC 7469 ...... & 3.3 & 5.4 & 8.6 & 41 & 54 & 56\\
\enddata
\tablenotetext{a}{In units of mJy.}
\end{deluxetable}

\begin{deluxetable}{cc r@{}c@{}l r@{}c@{}lc}
\tablecolumns{8}
\tablewidth{0pc}
\tablecaption{Coefficients for seeing correction in equation
\ref{seecor_eq} \label{seecor}}
\tablehead{
 Object & Band & \multicolumn{3}{c}{$b$} & \multicolumn{3}{c}{$c$}
}
\startdata
NGC 5548 ...... & $B$ & $-$0&.&025 & 0&.&058 \\
                & $V$ & 0&.&001 & 0&.&095 \\
                & $H$ & 0&.&23 & 0&.&28 \\
                & $K$ & 0&.&41 & &-&\\
NGC 4051 ...... & $V$ & $-$0&.&02 & 0&.&066 \\
                & $K$ & 0&.&006 & 0&.&39 \\
NGC 7469 ...... & $U$ & 0&.&058 & &-&\\
                & $B$ & $-$0&.&007 & 0&.&066\\
                & $V$ & 0&.&066 & 0&.&075\\
                & $J$ & 0&.&77 & &-&\\
                & $H$ & 0&.&54 & 0&.&81\\
                & $K$ & 0&.&67 & 0&.&55\\
\enddata
\tablecomments{No seeing correction was attempted in the cases of $BJH$ 
for NGC 4051 and $BVJHK$ for NGC 3227, because of larger photometric 
errors.}
\end{deluxetable}

\begin{deluxetable}{ccc rrc c rrc}
\tablecolumns{10}
\tablewidth{0pc}
\tablecaption{Sampling statistics for the $V$ and $K$ bands \label{smpstat}}
\tablehead{
 & & & \multicolumn{3}{c}{$V$ Band} && \multicolumn{3}{c}{$K$ Band}\\
\cline{4-6}\cline{8-10}
Object & Span & $\Delta T_{\rm max}$ & $N$ & $\Delta T_{\rm ave}$ & $\Delta T_{\rm med}$  &&  
                                       $N$ & $\Delta T_{\rm ave}$ & $\Delta T_{\rm med}$ 
}
\startdata                                   
NGC 5548 ...... & 893 & 110 & 101 &  8.9 & 3  &&  95 &  9.5 & 4\\
NGC 4051 ...... & 915 & 112 &  84 & 11.0 & 5  &&  89 & 10.4 & 5\\
NGC 3227 ...... & 589 & 151 &  40 & 15.1 & 4  &&  40 & 15.1 & 4\\
NGC 7469 ...... & 748 & 154 &  88 &  8.6 & 3  &&  86 &  8.8 & 3\\
\enddata
\tablecomments{$N$ is the number of data points, and others are
various time intervals in units of days.}
\end{deluxetable}

\begin{deluxetable}{cccccccccc}
\tablecolumns{10}
\tablewidth{0pc}
\tablecaption{Variability statistics in the $V$ and $K$ bands \label{varstat}}
\tablehead{
& \multicolumn{4}{c}{$V$ Band} && \multicolumn{4}{c}{$K$ Band}\\
\cline{2-5}\cline{7-10}
Object & $\left<f\right>$ & $\sigma$ & $F_{\rm var}$ & $R_{\rm max}$ && 
$\left<f\right>$ & $\sigma$ & $F_{\rm var}$ & $R_{\rm max}$
}
\startdata
NGC 5548 ...... & 2.6 & 0.9 & 0.32 & 3.2 && 16 & 4 & 0.26 & 2.9 \\
NGC 4051 ...... & 3.6 & 0.6 & 0.16 & 2.0 && 31 & 7 & 0.22 & 2.5 \\
NGC 3227 ...... & 6.6 & 0.4 & 0.06 & 1.3 && 18 & 3 & 0.18 & 2.0 \\
NGC 7469 ...... & 5.8 & 0.6 & 0.10 & 1.8 && 41 & 3 & 0.06 & 1.2 \\
\enddata
\tablecomments{$\left<f\right>$ is the mean flux, and $\sigma$ is the rms variation in units 
of mJy. $F_{\rm var}$ is the normalized variability amplitude defined in 
equation \ref{mf_eq}, and $R_{\rm max}$ is the ratio of maximum to minimum 
fluxes defined as $f_{{\rm max}}/f_{{\rm min}}$.}
\end{deluxetable}

\begin{deluxetable}{cccr r@{}lc c r@{}lc}
\tablecolumns{15}
\tablewidth{0pc}
\tablecaption{Cross-correlation results from the observed light curves \label{tp_VK}}
\tablehead{
&  &  &  &  \multicolumn{3}{c}{BI Method} && \multicolumn{3}{c}{ES Method}\\
\cline{5-7}\cline{9-11}
Object & Subset & MJD & $N$ & \multicolumn{2}{c}{$\tau_{\rm peak}$} & $r_{\rm max}$ && \multicolumn{2}{c}{$\tau_{\rm peak}$} & $r_{\rm max}$  
}
\startdata
NGC 5548 ...... & All data       & $51992-52885$ & 101 &  45&.1 & 0.909 &&  49&.2 & 0.920\\
                & Year 2001      & $51992-52163$ &  36 &  47&.8 & 0.962 &&  48&.0 & 0.961\\
                & Year 2002-2003 & $52265-52823$ &  52 &  49&.5 & 0.968 &&  48&.3 & 0.947\\
                & Year 2003      & $52622-52885$ &  42 &  57&.0 & 0.952 &&  53&.1 & 0.946\\
NGC 4051 ...... & All data       & $51917-52819$ &  84 &  $>$5&0\tablenotemark{a}&   -   &&  $>$5&0\tablenotemark{a}&   -  \\
                & Year 2001      & $51917-52117$ &  28 &   9&.0 & 0.347 &&  10&.3 & 0.479\\
                & Year 2002      & $52229-52488$ &  30 &  11&.3 & 0.894 &&  11&.4 & 0.886\\
                & Year 2003      & $52599-52819$ &  26 &  25&.9 & 0.804 &&  17&.7 & 0.779\\
NGC 3227 ...... & All data       & $52226-52815$ &  40 &  10&.9 & 0.640 && $<$8&\tablenotemark{b}  &   -  \\
                & Year 2002      & $52226-52423$ &  12 &  16&.7 & 0.701 &&  18&.9 & 0.772\\
                & Year 2003      & $52574-52815$ &  28 &   6&.7 & 0.714 && $<$5&\tablenotemark{b}  &   -  \\
NGC 7469 ...... & All data       & $52075-52647$ &  88 &  87&.7 & 0.716 &&  86&.0 & 0.606\\
                & Year 2001      & $52075-52265$ &  53 &  88&.8 & 0.798 &&  87&.9 & 0.822\\
                & Year 2002      & $52419-52647$ &  30 &  52&.9 & 0.846 &&  55&.4 & 0.826\\
\enddata
\tablecomments{$N$ is the number of data points. $\tau_{\rm peak}$ is the position 
at which the CCF peak occurs, and $r_{\rm max}$ is the peak value of CCF.}
\tablenotetext{a}{The lower bound of $\tau_{\rm peak}$ above which the CCF 
shows a very broad maximum.}
\tablenotetext{b}{The upper bound of $\tau_{\rm peak}$ below which the CCF 
shows a plateau-like maximum with no clear peak.}
\end{deluxetable}

\begin{deluxetable}{lcccclcc}
\tablecolumns{8}
\tablewidth{0pc}
\tablecaption{Structure functions from the observed light curves \label{sf}}
\tablehead{
& \multicolumn{3}{c}{$V$ Band} && \multicolumn{3}{c}{$K$ Band} \nl
\cline{2-4} \cline{6-8}
Object & $S_{0}$  & $\beta$ & $\tau_{\rm max}$  && 
$S_{0}$ & $\beta$ & $\tau_{\rm max}$
}
\startdata
NGC 5548 ...... & $1.8\times10^{-3}$ & 1.46 & 60 && $1.7\times10^{-3}$ & $2.0$ & 60 \nl
NGC 4051 ...... & $2.4\times10^{-2}$ & 0.71 & 30 && $7.9\times10^{-2}$ & $1.5$ & 30 \nl
NGC 3227 ...... & $2.1\times10^{-2}$ & 0.91 & 20 && $4.0\times10^{-2}$ & $1.1$ & 50 \nl
NGC 7469 ...... & $3.2\times10^{-2}$ & 0.82 & 30 && $1.6\times10^{-2}$ & $1.6$ & 50 \nl
\enddata
\tablecomments{$S_0$ is the fiducial flux in units of $({\rm mJy})^2$ at time 
   interval of one day, $\beta$ is the power-law index of structure function,
   and $\tau_{\rm max}$ is the upper bound of days for power-law fit.}
\end{deluxetable}

\begin{deluxetable}{cc ccc cr@{}l cr@{}l}
\tablecolumns{11}
\rotate
\tablewidth{0pc}
\tablecaption{Cross-correlation results from the simulated light curves \label{lag_VK}}
\tablehead{
Object & Subset & $M_{V}$ & Magnitude Range &  & \multicolumn{3}{c}{Lag Time (BI)} & \multicolumn{3}{c}{Lag Time (ES)}  
}
\startdata
NGC 5548 ...... & All data       & $-19.19$ & $[-18.39,\; -19.65\; ]$ &  & &$45.1$&${}_{-3.4}^{+1.5}$     & & $48.7$&${}_{-4.1}^{+4.6}$\\
                & Year 2001      & $-19.39$ & $[-18.88,\; -19.65\; ]$ &  & &$47.8$&${}_{-1.2}^{+1.7}$     & & $48.0$&${}_{-2.3}^{+2.7}$\\
                & Year 2002-2003 & $-19.04$ & $[-18.39,\; -19.60\; ]$ &  & &$47.3$&${}_{-5.6}^{+3.9}$     & & $46.7$&${}_{-5.3}^{+4.8}$\\
                & Year 2003      & $-19.18$ & $[-18.39,\; -19.60\; ]$ &  & &$57.0$&${}_{-6.2}^{+1.2}$     & & $52.9$&${}_{-5.8}^{+3.0}$\\
NGC 4051 ...... & All data       & $-15.65$ & $[-15.27,\; -16.02\; ]$ &  & &{\bf $24.1$}&{\bf ${}_{-9.2}^{+11.1}$}    & &{\bf $21.0$}&{\bf ${}_{-4.8}^{+8.4}$}\\
                & Year 2001      & $-15.51$ & $[-15.27,\; -15.76\; ]$ &  & &{\bf $10.2$}&{\bf ${}_{-2.2}^{+16.4}$}    & &{\bf $10.9$}&{\bf ${}_{-3.7}^{+4.2}$}\\
                & Year 2002      & $-15.66$ & $[-15.38,\; -16.02\; ]$ &  & &$11.0$&${}_{-2.9}^{+4.4}$     & & $10.8$&${}_{-4.0}^{+5.5}$\\
                & Year 2003      & $-15.77$ & $[-15.56,\; -16.02\; ]$ &  & &$22.8$&${}_{-5.9}^{+3.1}$     & &{\bf $18.1$}&{\bf ${}_{-4.4}^{+6.2}$}\\
NGC 3227 ...... & All data       & $-17.35$ & $[-17.15,\; -17.47\; ]$ &  & &{\bf $6.8$}&{\bf ${}_{-4.6}^{+4.1}$}      & &   -  &\\
                & Year 2002      & $-17.35$ & $[-17.23,\; -17.43\; ]$ &  & &{\bf $16.3$}&{\bf ${}_{-4.4}^{+6.0}$}     & &{\bf $20.2$}&{\bf ${}_{-5.7}^{+5.2}$}\\
                & Year 2003      & $-17.35$ & $[-17.15,\; -17.47\; ]$ &  & &$6.0$&${}_{-5.4}^{+4.8}$      & &   -   &\\
NGC 7469 ...... & All data       & $-19.78$ & $[-19.40,\; -20.00\; ]$ &  & &{\bf $87.6$}&{\bf ${}_{-14.3}^{+16.7}$}   & &{\bf $85.4$}&{\bf ${}_{-24.3}^{+26.7}$}\\
                & Year 2001      & $-19.82$ & $[-19.60,\; -20.00\; ]$ &  & &$88.0$&${}_{-8.4}^{+17.8}$    & & $86.9$&${}_{-14.1}^{+17.5}$\\
                & Year 2002      & $-19.71$ & $[-19.39,\; -19.96\; ]$ &  & &$77.7$&${}_{-25.5}^{+21.7}$   & &{\bf $64.9$}&{\bf ${}_{-12.8}^{+29.4}$}\\
\enddata
\tablecomments{$M_V$ is the absolute $V$-band magnitude averaged over the magnitude range observed.
The lag time is a median of $\tau_{\rm peak}$ in the CCPD.
For NGC 3227, when the CCPD results in too broad a distribution, reflecting the 
plateau feature in the CCF (Figure \ref{ccf_ngc3227}),
the lag time is given no value in the blank column.
}
\end{deluxetable}

\end{document}